\RequirePackage{fix-cm}
\documentclass{svjour3}                 
\smartqed  
\usepackage{graphicx,graphics,amssymb}
\usepackage{bm}
\usepackage{color}
\usepackage{mathtools}
\usepackage{natbib}
\usepackage{caption}
\usepackage{subcaption}
\usepackage{float}

\journalname{Transport in Porous Media}
\begin{document}

\title{A Primer on the Dynamical Systems Approach to Transport in Porous Media
}

%


\titlerunning{A Primer on the Dynamical Systems Approach}        

\author{Guy Metcalfe         \and
        Daniel Lester        \and
        Michael Trefry
}


\institute{G. Metcalfe \at
              Swinburne University of Technology, Melbourne Australia \\
              \email{gmetcalfe@swin.edu.au}           
           \and
           D. Lester \at
              School of Engineering, RMIT University, Melbourne Australia \\
              \email{daniel.lester@rmit.edu.au}
            \and
            M. Trefry \at
              Independent Researcher, Perth Australia \\
              \email{michaelgtrefry@gmail.com}
}

\date{\today}

\maketitle

\begin{abstract}
Historically, the dominant conceptual paradigm of porous media flow, solute mixing and transport was based on steady two-dimensional flows in heterogeneous porous media. Although it is now well recognised that novel transport phenomena can arise in unsteady and/or three-dimensional flows at both the pore- or Darcy-scales, appropriate methods for analysis and understanding of these more complex flows have not been widely employed. In this primer we advocate for methods borrowed from dynamical systems (chaos) theory, which aim to uncover the \emph{Lagrangian kinematics} of these flows: namely how fluid particle trajectories (which form a dynamical system) are organized and interact and the associated impacts on solute transport and mixing. This dynamical systems approach to transport is inherently Lagrangian, and the Lagrangian kinematics form Lagrangian coherent structures (LCSs), special sets of trajectories that divide the Lagrangian frame into chaotic mixing regions, poorly mixing hold-up regions (and in some cases non-mixing ``islands'') and the transport barriers that organise these regions. Hence the dynamical systems approach provides insights into flows that may exhibit chaotic, regular (non-chaotic) or mixed Lagrangian kinematics, and also into how LCSs organize solute transport and mixing. Novel experimental methods are only recently permitting visualization of LCSs are in porous media flows. In this primer we review the dynamical systems approach to porous media flow and transport and connect the associated tools and techniques with the latest research findings from pore to Darcy scales. This primer provides an introduction to the methods and tools of dynamical systems theory. Once familiar with these approaches, porous media researchers will be better positioned to know when to expect complex Lagrangian kinematics, how to uncover and understand LCSs and their impacts on solute transport, and how to exploit these dynamics to control solute transport in porous media flows. 
\end{abstract}


\section{Introduction}
\label{sec:intro}

Since the pioneering work of \citet{Darcy:1856aa} regarding the quantitative relationship between fluid pressure gradient and flux through soil columns, interest in the fluid dynamics of porous media systems has grown rapidly. These dynamics govern the transport, reaction and biological activity of solutes, colloids and microorganisms in a wide range of porous materials ranging from geological media to biological tissues and engineered structures, and now underpin many vital industries, including biotechnology, energy, agriculture and water supply. All porous media are characterised by complex pore-scale architectures and many also exhibit significant heterogeneities from pore scales to much larger (e.g. regional) length scales, resulting in complex flow and transport dynamics not resolved in Darcy's original work. Due to these inherent complexities, the conceptual progress of porous media science has not been smooth. Progress has relied on irregular advances in technology that have enabled greater precision and resolution in observation, which have then provided the data that seeded the next conceptual advance. Today, researchers have unparalleled access to pore-scale data of fluid flow, three-dimensional (3D) geometry and surface chemistry, yielding an almost overwhelmingly detailed and rich picture of the process complexity inherent in porous media. The next conceptual advance is imminent.

\begin{figure}[tb]
\centering
\includegraphics[width=0.95\columnwidth]{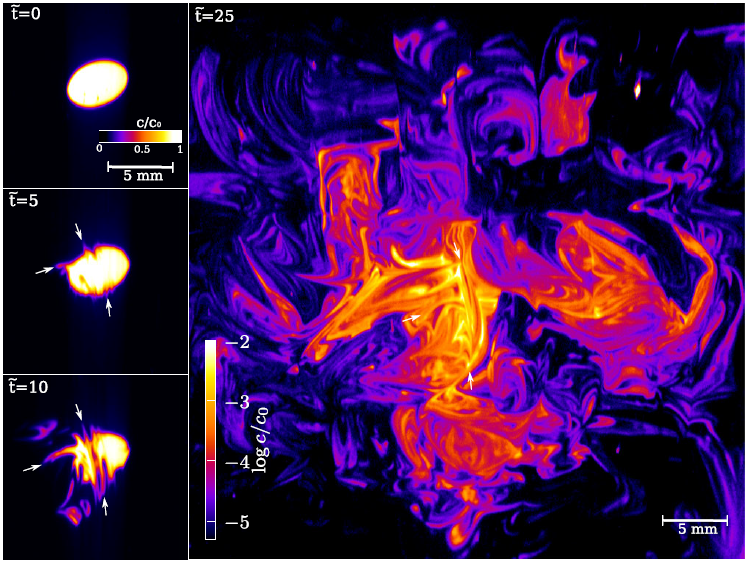}
\caption{Heterogeneous and anisotropic dispersion of a solute plume (imaged near the injection point in a plane oriented normal to the mean flow direction) during the pull phase of a push-pull experiment through randomly packed, index-matched glass beads. The initially ellipsoidal dye cross section at the beginning of the pull phase (given by $\tilde{t} = 0$) diffuses preferentially along directions of high fluid deformation
(indicated by white arrows), leading to a highly striated dye signature at later times.  Adapted from \citet{Heyman:2021aa}.}
\label{fig:pushpullexperiment}
\end{figure}

Figure~\ref{fig:pushpullexperiment} presents an example of the kinds of complex experimental data now available and awaiting interpretation. The figure shows the results of a standard push-pull experiment at low Reynolds number where the concentration distribution of a fully developed solute plume near the injection point is shown at different times $\tilde{t}$ during the pull phase. The width of the injected plume shown at $\tilde{t}=0$ has similar dimensions to those of the glass beads used to form the randomly packed porous medium. In the absence of molecular diffusion, reversibility of the push-pull flow would result in the collapse of the solute plume during the pull phase back to the original injection profile shown at $\tilde{t}=0$. Hence the solute plume shown in Figure~\ref{fig:pushpullexperiment} represents a residual ``diffusive signature'' resulting from the irreversible interactions between fluid deformation history (noting that net fluid deformation is zero) and solute diffusion. After completion of the pull phase, at $\tilde{t} = 25$, this diffusive signature is observed as an unexpectedly extensive and striated distribution. This distribution indicates the presence of profound anisotropic fluid transport phenomena at the pore scale which are unseen in less-resolved investigations, and cannot be described via a conventional solute dispersion framework. Novel tools and techniques are required to understand, characterise and quantify such transport phenomena.

\begin{figure}[tb]
\centering
\includegraphics[width=0.9\columnwidth]{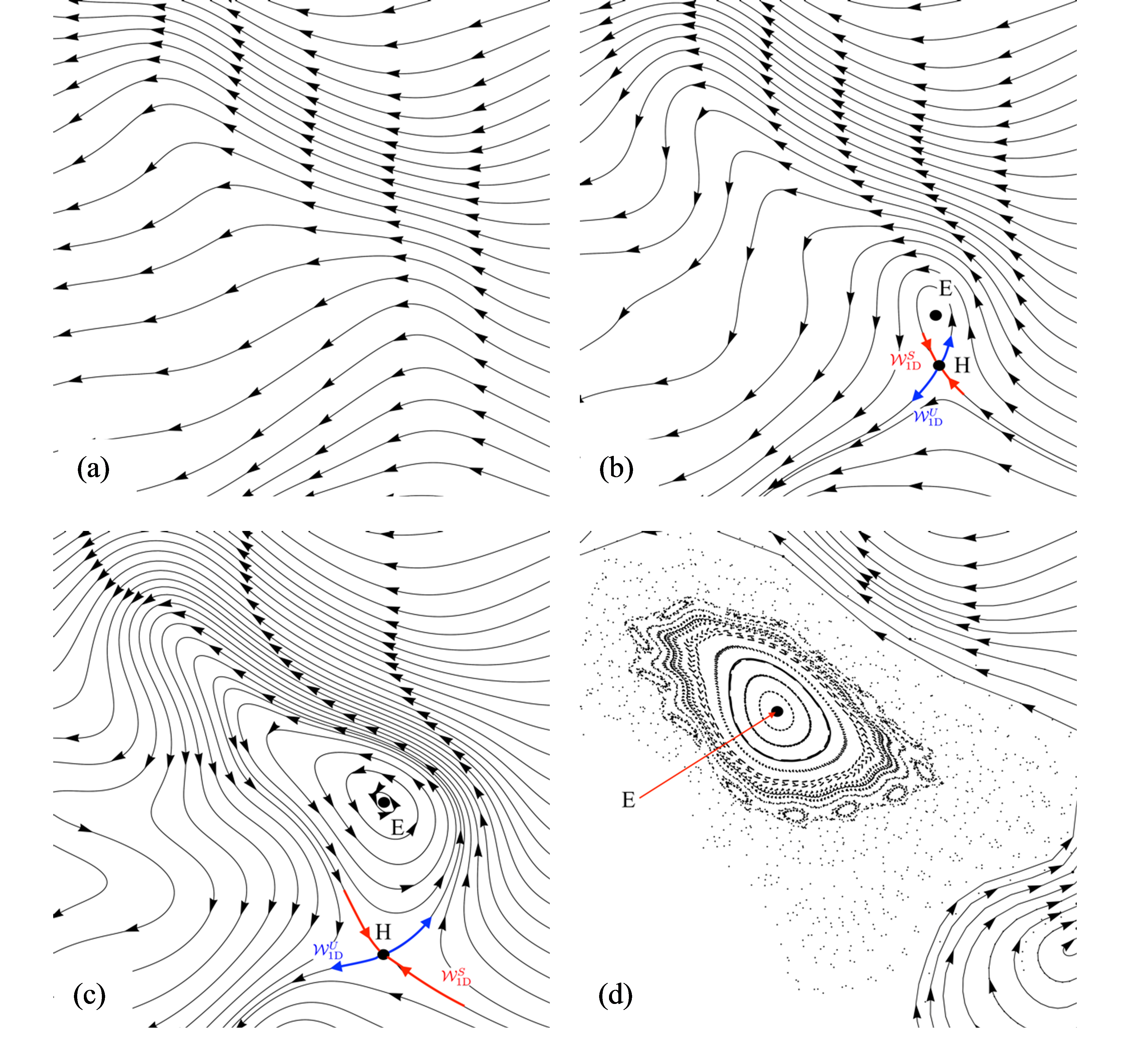}
\caption{Lagrangian coherent structures (LCSs) in an unsteady heterogeneous, poroelastic Darcy flow as a function of increasing medium compressibility $\mathcal{C}$. 
With increasing compressibility $\mathcal{C}$, the transport structure of the Darcy flow bifurcates from open trajectories (a) to a closed region (b,c) with elliptic (E) and hyperbolic (H) periodic points and associated stable $(\mathcal{W}_{1D}^{s})$ and unstable $(\mathcal{W}_{1D}^{u})$ manifolds. (d) At large $\mathcal{C}$ the hyperbolic point (H) bifurcates to a chaotic mixing region, while part of the closed region around the elliptic point (E) persists. Adapted from \citet{Wu:2019aa}.}
\label{fig:varyC}
\end{figure}

At the same time, it has also been discovered that quite unexpected transport phenomena can present in naturally occurring flows at the Darcy scale, i.e. well beyond the pore scale. Using a conventional poroelastic Darcian flow model, \cite{Wu:2019aa} showed that simply by increasing the compressibility $\mathcal{C}$ of the porous medium a range of transport structures of increasing complexity can be generated. Figure~\ref{fig:varyC} depicts the motion of tracer particles (black points and lines) at integer multiples of the tidal forcing period $P$ in a two-dimensional (2D) model of a compressible heterogeneous aquifer subject to both tidal boundary forcing and a background regional flow. As shown in Figure~\ref{fig:varyC}a, for small values of $\mathcal{C}$ tracer particles sampled at integer multiples of $P$ move along curved paths from inland (right) toward the coastal boundary (left). With increasing compressibility (Figure~\ref{fig:varyC}b, c), elliptic (E) and hyperbolic (H) periodic points arise in the aquifer. The local fluid motion (over one flow period) around the elliptic point (E) is rotational, while fluid stretching and compression (indicated by the stable (red) and unstable (blue) trajectories) occurs local to the hyperbolic point (H). Associated with this bifurcation in the transport structure of the flow is the appearance of a closed region (where non-diffusive tracer particles are trapped indefinitely) around the elliptic point, indicating a qualitative change in the aquifer transport dynamics. Above a critical compressibility (Figure~\ref{fig:varyC}d), the hyperbolic point (H) and the associated open region bifurcates into a chaotic region (indicated by the scattered points) and a regular region, while the closed region persists as a stable ``island'' with the elliptic point (E) at its centre. This chaotic region exhibits complex solute transport dynamics, including stochastic residence time signatures and accelerated mixing.  The above transport structures are not easy to predict, detect or understand using standard techniques, yet they exert significant control on solute transport.

From the above examples it is clear that conventional Darcy- and pore-scale tools and techniques (that are so well suited to many practical problems) are unlikely to be of much use in understanding and quantifying the transport and mixing dynamics of these more complex flows. A conceptual advance is required to push the science further, and this requires consideration of the \emph{Lagrangian kinematics} of these flows to elucidate how fluid particle trajectories are organised by the underlying transport structure. This underlying transport structure governs processes ranging from solute mixing and pollutant dispersion, to chemical reactions and biological activity. The natural language of these kinematics is couched in terms of dynamical systems (chaos) theory, and in this paper we introduce the key concepts in this field required to understand solute transport in porous media using a dynamical systems approach. It is important to note that the dynamical systems approach to understanding the Lagrangian kinematics of porous media flows is not limited to understanding chaotic particle trajectories (i.e. ``chaotic advection''), but, as shall be shown, this approach is also useful in understanding transport at a deeper level in regular (non-chaotic) systems, or systems such as shown in Figure~\ref{fig:varyC}(d) that exhibit both regular and chaotic regions. This primer on the dynamical systems approach is deliberately pitched to provide an overview of the field rather than a detailed exposition, and the interested reader is directed to the referenced works for greater detail. As we will see in the following sections, dynamical systems theory brings with it a rich set of concepts and tools relevant to modern porous media flow, and it already has a (relatively short) history in the subject.

\begin{figure}[tb]
\centering
\includegraphics[width=0.9\columnwidth]{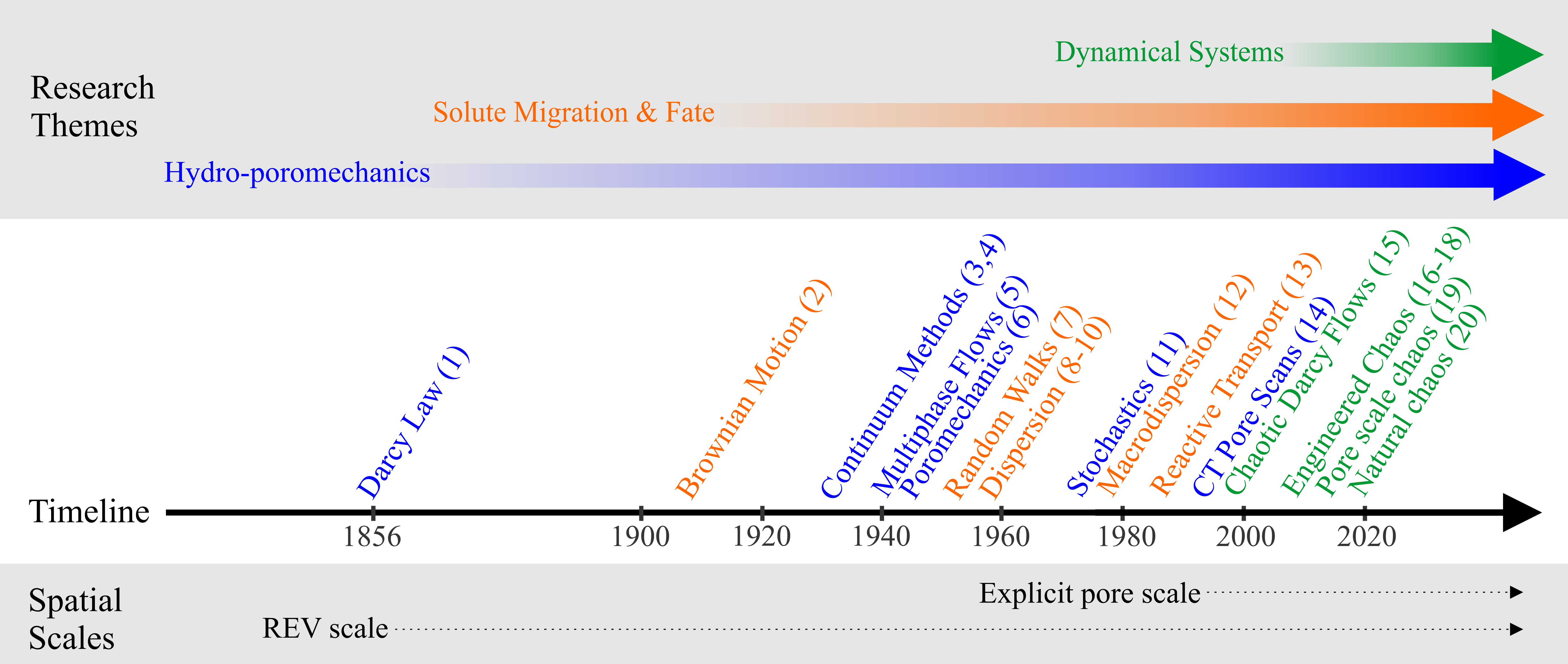}
\caption{Timeline of porous media research showing some key research themes and scales of investigation, and emphasizing a recent articles on the dynamical systems approach to understanding transport in porous media (green).\\\hspace{\textwidth}(1) \citet{Darcy:1856aa};(2) \citet{Einstein:1905aa}; (3) \citet{Lamb:1932aa}; (4) \citet{Polubarinova:1938aa}: see \citet{Zlotnik:2007aa}; (5) \citet{Leverett:1941aa}, see \citet{Parker:1989aa}; (6) \citet{Terzaghi:1943aa}; (7) \citet{Scheidegger:1954aa}; (8) \citet{Saffman:1959aa}; (9) \citet{Scheidegger:1961aa}; (10) \citet{Bear:1961aa,Bear:2021aa}; (11) \citet{Gelhar:1974aa}; (12) \citet{Gelhar:1979aa}; (13) \citet{LICHTNER:1985aa}; (14) \citet{Spanne:1994aa}; (15) \citet{weeks:1998aa}; (16) \citet{Metcalfe:2009aa}; (17)  \citet{Mays:2012}; (18) \citet{Cho:2019aa}; (19) \citet{Lester:2013aa}; (20) \citet{Trefry:2019aa}.}
\label{fig:timeline}
\end{figure}

Taking a broad (and deliberately highly selective) overview of porous media research, we may form a timeline of key concepts and breakthroughs in transport in porous media, depicted in Figure~\ref{fig:timeline}. The timeline shows the recent application of dynamical systems concepts to the field, in green. The dynamical systems entries are disproportionately over-represented in the timeline as each of the other research themes have accrued many thousands of articles over the decades while there are still only a handful of dynamical systems articles in the porous media literature. However we contend that the new tools and techniques provided by dynamical systems theory may facilitate rapid advances in our understanding of flow and transport in porous systems, for the following reasons.

It is now well recognized that several classes of porous media flows give rise to transport phenomena that are more complex than steady 2D flow, and so cannot be fully understood via conventional analysis~\citep{weeks:1998aa}. Select examples of complex steady Darcy flows include helical motion in steady anisotropic 3D Darcy flow~\citep{Bakker:2004aa,Cirpka:2012aa,Cirpka:2015aa, Chiogna:2014aa,Ye:2015aa}, steady 3D Darcy flow in an otherwise homogeneous porous medium with spheroidal inclusions~\citep{DiDato:2016aa,DiDato:2016ab,Jankovic:2003aa,Jankovic:2009aa}, while complex unsteady Darcy flows have been observed in both homogeneous~\citep{Bagtzoglou:2007aa,Metcalfe:2010aa,Trefry:2012aa,Mays:2012,Piscopo:2013aa,Sather:2022aa} and hetereogeneous~\citep{Trefry:2019aa,Cho_field_2014} media. Similarly, complex steady 3D flows have been shown to arise at the pore-scale in both open porous networks~\citep{Lester:2013ab,Lester:2014aa} and granular media~\citep{Kree:2017aa,Turuban:2019aa,souzy:2020aa,Heyman:2020aa}. Although the complexity of these flows arises via different mechanisms---ranging from anisotropy of the conductivity field to transient forcing to complexity of the pore-scale architecture---these flows are unified in that their fluid particle trajectories are organised in a more complicated manner than the lamellar streamlines of all steady 2D flows, and so the Lagrangian kinematics of these flows cannot be understood solely from the Eulerian velocity field. Similarly, solute mixing in all of these flows is characterised by highly striated, ramified solute distributions that cannot be fully understood from the Eulerian velocity field.

Although the Lagrangian frame has an extensive history of application in the study of transport in porous media, use of Lagrangian coordinates alone is insufficient to understand transport. Indeed, several of the studies outlined above have not explicitly resolved the Lagrangian kinematics of these flows, which is concerned with the organisation of fluid trajectories and the attendant deformation of fluid elements (see \citet{Trefry:2019aa} for an example of organised kinematics). In this primer, we advocate for a Lagrangian coordinate description of particle trajectories, but the important next step to understand is that particle trajectories (because they form a dynamical system) organize into coherent structures that can be understood by analyzing special sets of trajectories using tools developed in dynamical systems theory. Hence the dynamical systems theory approach to transport in porous media provides a convenient framework for detecting and classifying kinematic features in the Lagrangian frame that can control fluid mixing, segregation and discharge in porous media.  These kinds of transport phenomena describe complex Lagrangian structures (termed \emph{Lagrangian Coherent Structures} (LCSs)) which engender potentially profound impacts on solute migration and reaction \citep{Toroczkai:1998aa,Tel2005,Valocchi:2019aa}. Coupled with appropriate stochastic methods (such as random walks and non-local transport theories), the Lagrangian kinematics can be used to develop quantitative predictions of solute transport, mixing and reaction that resolve the complex dynamics observed in Figures~\ref{fig:pushpullexperiment} and \ref{fig:varyC}. 

One the first authors to use a dynamical systems approach to understand flow and transport in porous media was \citet{Sposito:1994aa}, who recognised that steady Darcy flow in heterogeneous, isotropic porous media imposes strong constraints on the Lagrangian kinematics, precluding chaotic motion. Although this class of porous medium flow is not complex in the manner described above, only explicit consideration of the Lagrangian kinematics uncovers these properties. Conversely, early conceptual work by \citet{Ottino:1989aa,Jones:1988aa}, \citet{Sposito:2001aa,Sposito:2006aa} and others \citep{Lester:2009ab,Lester:2010aa,Metcalfe:2010aa,Trefry:2012aa,Mays:2012,Lester:2016aa} has led to the prediction, observation and engineering of complex Lagrangian kinematics in saturated porous media at the Darcy \citep{Zhang:2009aa,Metcalfe:2010ab} and field \citep{Cho:2019aa} scales. Complex Lagrangian kinematics have also been predicted to occur in natural environments \citep{Trefry:2019aa,Wu:2019aa}, e.g. compressible Darcian systems where engineered pumping activity is absent. Dynamical systems concepts have also been used to link the topology of pore-scale architectures with pore-scale solute mixing and macroscopic transport phenomena~\citep{Lester:2013aa,Lester:2014aa,Lester:2016ad}, resulting in fundamental upscaling behaviours conditioned on the details of pore-scale Lagrangian kinematics. These mechanisms have recently been confirmed via direct experimental observations of pore-scale mixing~\citep{souzy:2020aa,Heyman:2021aa,Heyman:2020aa}, leading to the highly striated solute distributions such as those shown in Figure~\ref{fig:pushpullexperiment}. Dynamical systems approaches also shed new light on other fluid-borne phenomena \citep{Tel2005} and thus Lagrangian kinematics underpin the transport, spreading, clustering and reactivity of solutes, colloids and microorganisms in porous media. Hence it is natural to seek new learnings from the dynamical systems approach to transport in porous media.

But what is dynamical systems theory, what can it tell us about the fundamental characteristics of flow, and how can it be applied to porous media systems? We attempt to answer these questions in the remainder of this paper. Our approach is to provide clear and concise definitions of key dynamical terms and concepts, illustrated by examples from the recent research literature, so that the reader is armed with the basic tools and knowledge to consider future applications of this powerful theory in the context of porous media.

\section{What is the Dynamical Systems Approach to Fluid Transport?}
\label{sec:whatischaos}

The dynamical systems approach to fluid transport, often termed ``chaotic advection'' (although the concepts may also be applied to non-chaotic systems), is the science at the intersection of fluid mechanics and nonlinear dynamical systems~\citep{Aref_frontiers_2017}. In this case the phase space (coordinates of interest) of the dynamical system is the physical space that the fluid occupies, whether at the pore scale or the Darcy scale, and either in a two-dimensional approximation or the full three-dimensional space. 

Hence chaotic advection is concerned with the trajectories of the dynamical system given by the collective motion of all fluid particles (where ``particle'' refers to a conceptual massless tracer particle) given by the Lagrangian kinematics of the flow.  The main point in this section is that particular structures in the flow --- collections of points, lines, surfaces --- organize the Lagrangian transport of the \emph{entire} flow, and transport can only be truly understood by uncovering the Lagrangian kinematics.

These Lagrangian coherent structures (LCS), such as those shown in Figure~\ref{fig:varyC}, form the ``skeleton'' of the flow that organises global transport~\citep{MacKay:1994}, even in the presence of random processes such as diffusion. And to understand or design transport in porous media flows (be it of heat, of chemical or biological species, of matter, in fact, of all interactions mediated by the flow) it is important and useful to find, classify, and elucidate these LCS. Broadly interpreted these Lagrangian structures consist of attracting regions, repelling regions, hold up regions, and the distinguished pathways that connect one critical region to another. 

In fluids, including porous media flows from pore to regional scales, the Lagrangian trajectories are the trajectories of fluid particles. From its initial location the trajectory of every fluid particle is given by the kinematic equation
\begin{align}
\label{eq:kinematic}
\frac{dx}{dt} = V_x && \frac{dy}{dt} = V_y && x(t=0)=x_0 && y(t=0)=y_0
\end{align}
here written in 2D with the fluid particle location $(x,y)$ moved passively by the fluid velocity vector field $(V_x, V_y)$.  Deceptively simple, the kinematic equation can be taken as the elementary definition of velocity or as defining a nonlinear dynamical system in which a given velocity field generates the Lagrangian trajectories of the fluid tracer particles. Hence these trajectories provide a coordinate transform between the Eulerian ($x,y$) and Lagrangian ($x_0$, $y_0$) frames. In keeping with the idea of building from simplest concepts first, the kinematic equation (\ref{eq:kinematic}) has no diffusion or local dispersion.

Throughout this primer we focus mainly on the Lagrangian kinematics (which are purely advective), but we note that these kinematics govern fluid deformation (i.e. the stretching, shearing, and folding of fluid elements)~\citep{Lester:2018aa}, and solute transport in the presence of diffusion or local dispersion is controlled by the deformation history of fluid elements~\citep{Villermaux:2019aa}. Hence such additional physics can be readily incorporated, but knowledge and quantification of the Lagrangian kinematics is essential to correctly condition the random fluid stretching processes that govern the advection-diffusion process.

\begin{figure}[tbp]
\centering
\begin{subfigure}[b]{0.27\textwidth}
    \centering
    \includegraphics[width=\textwidth]{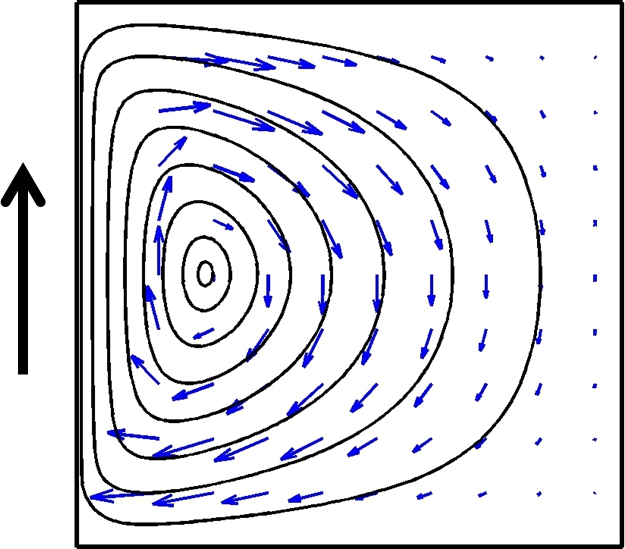}\label{LidDrivenCavity_a}
    \caption{First half period}
    \label{fig:1sthalfperiod}
\end{subfigure}
\begin{subfigure}[b]{0.27\textwidth}
    \centering
    \includegraphics[width=\textwidth]{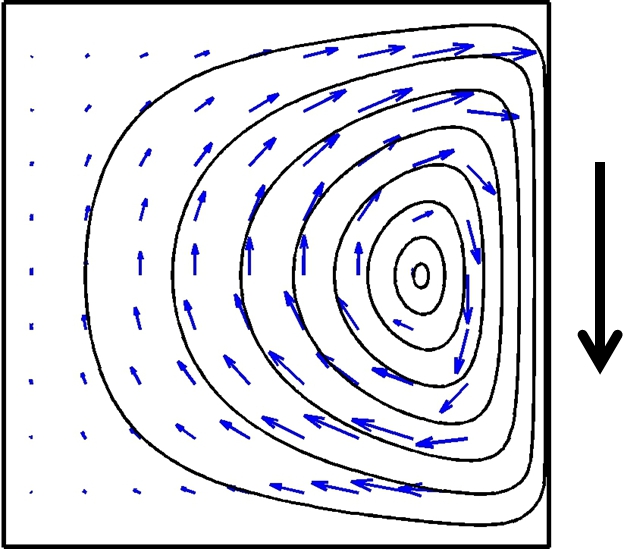}\label{LidDrivenCavity_b}
    \caption{Second half period}
    \label{fig:2ndhalfperiod}
\end{subfigure}
\begin{subfigure}[b]{0.23\textwidth}
    \centering
    \includegraphics[width=\textwidth]{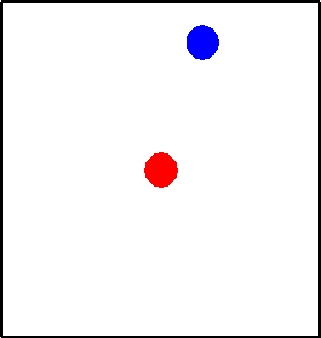}
    \caption{Initial state}
    \label{fig:IllustrateChaosparta}
\end{subfigure}
\begin{subfigure}[b]{0.25\textwidth}
    \centering
    \includegraphics[width=\textwidth]{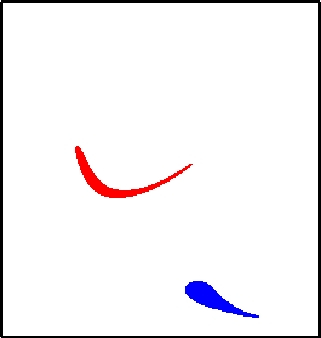}
    \caption{After 1 period}
    \label{fig:IllustrateChaospartb}
\end{subfigure}
\begin{subfigure}[b]{0.25\textwidth}
    \centering
    \includegraphics[width=\textwidth]{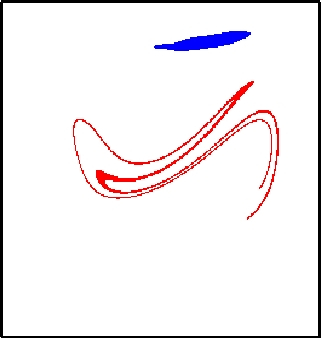}
    \caption{After 3 periods}
    \label{fig:IllustrateChaospartc}
\end{subfigure}
\begin{subfigure}[b]{0.25\textwidth}
    \centering
    \includegraphics[width=\textwidth]{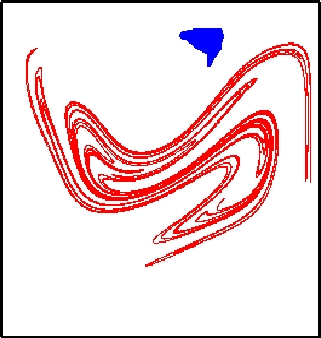}
    \caption{After 6 periods}
    \label{fig:IllustrateChaospartd}
\end{subfigure}
\begin{subfigure}[b]{0.4\textwidth}
    \centering
    \includegraphics[width=\textwidth]{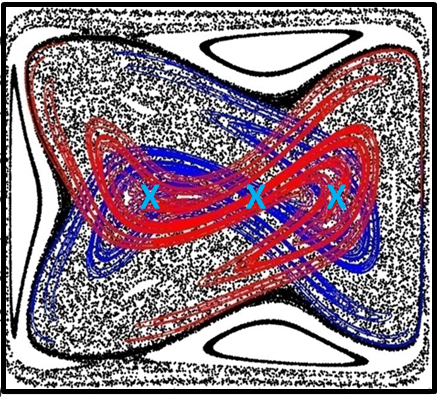}
    \caption{Typical Lagrangian flow structure.}
    \label{fig:OrderWithinChaos}
\end{subfigure}
\caption{Time-periodic 2D cavity flow driven by translating walls during the (a) first and (b) second half periods.  Blue arrows and closed curves indicate velocity vectors and streamlines. (c--f) Evolution of blue and red tracer blobs.  Points within the red blob are attracted and repelled from hyperbolic points (shown in (g) by cyan crosses) that create chaotic, spreading trajectories. The blue blob is in a hold-up/island region, experiencing little deformation and displacement.  (g) Typical hidden Lagrangian flow structure.  Black dots are the stroboscopic map of tracer particles showing well-mixed and island regions.  Blue (red) lines are attracting (repelling) manifolds of the central hyperbolic points. Adapted from~\protect\cite{Speetjens_3D_2021}.
}
\label{fig:2dPeriodicExample}
\end{figure}

Equation (\ref{eq:kinematic}) is deceptive in that it can generate complex trajectories even for simple velocity fields.  An example will illustrate a simple case and expose selected hidden LCS. Referring to Figure~\ref{fig:2dPeriodicExample}, our example is a 2D, time-periodic Stokes flow in a box of fluid.  This is amongst the simplest settings in which to observe LCS.  One sidewall steadily slides for half the period and drives a simple vortex flow as in Figure~\ref{fig:1sthalfperiod}; during the other half period, the opposing wall steadily slides in the opposite direction (Figure~\ref{fig:2ndhalfperiod}). A video of a similar experiment is also available~\citep{manifold_pileup_youtube}.

In Figure~\ref{fig:IllustrateChaosparta} two blobs of fluid are marked red or blue, and Figures~\ref{fig:IllustrateChaospartb}--\ref{fig:IllustrateChaospartd} show the advection-only evolution of the blobs after the indicated number of periods.  The evolution of the two blobs is markedly different because their deformation is being controlled by different types of LCS.  The blue blob was deliberately put into a hold-up region and undergoes a shear-like deformation that yields (at most) linear stretching, and every three periods it returns close to its initial condition. In 2D, time-periodic flows material starting in a hold-up region is truly cut off from the rest of the flow; these regions are called elliptic islands because the material inside the island rotates around an elliptic point (as exhibited in Figure~\ref{fig:varyC}(c,d)), and the net deformation around these periodic points is a simple rotation.  In 3D or non-periodic flows hold-up regions may not be completely cut off from the rest of the flow, but are still delineated by surfaces of minimal flux across them that keeps material close for long periods of time.

The red blob, on the other hand, is rapidly deformed into a highly ramified and extended filament.  The red blob was deliberately placed near a hyperbolic point (shown in Figure \ref{fig:2dPeriodicExample}(g)), where the net fluid deformation over one flow period consists of repelling (stretching) in two directions and attracting (compressing) in the two orthogonal directions (as shown in Figure~\ref{fig:varyC}(c,d))). In this simple example these hyperbolic points encapsulate both the attracting and repelling LCS; moreover, in this case the distinguished pathway between attracting and repelling regions begins and ends at the same point, which is denoted a homoclinic connection.  In general in 2D periodic flows heteroclinic connections, where the repelling and attracting points are different, are typical.  Here the hyperbolic point and its homoclinic connection strongly deforms the red blob through repeated stretching and folding.  This in turn generates exponential growth of the red blob's interfacial area, even though the blob remains confined inside a finite volume of fluid. 

That is why the shape of the red blob becomes more and more ramified with smaller and smaller intervals of unmarked fluid between the red filaments.  Even for fluid particles that are initially arbitrarily close together it cannot be predicted how far apart they will end up.  This destruction of long term predictability of the relative location of fluid particles is why such flows are called {\em chaotic}, even though the underlying velocity field is deterministic \citep{Ottino:1989aa}. 

That the chaotic part of a flow generates small-scale structure in the spatial distribution of material (the red blob) is why chaotic advection is associated with mixing. The chaotic part of the flow quickly evolves any smooth initial distribution into a complex pattern of filaments or sheets, depending on the dimensionality of the system, which converges at an exponential rate to a fractal structure. When the characteristic length scale of the blob becomes small enough, this structure is then smoothed by diffusion, manifesting as accelerated mixing. This rapid reduction of length scale and these augmented transport characteristics are purely effects of the simple kinematic equation (\ref{eq:kinematic}).

Figure~\ref{fig:OrderWithinChaos} illustrates several ways to bring out the Lagrangian structure that are commonly used in the dynamical systems literature.  The black dots are a ``stroboscopic map'' of tracer particles similar to that used in Figure~\ref{fig:varyC}. Take a small set of initial conditions (a line of particles at $x = 0.5$ in Figure~\ref{fig:OrderWithinChaos}), and for each fluid particle evolve its trajectory and place a dot to mark the particle's location after every period.  Continue this process for many periods and the resulting collection of dots defines a \emph{stroboscopic map} that visualises the long term Lagrangian structure.  For instance, in Figure~\ref{fig:OrderWithinChaos} the stroboscopic map clearly partitions the fluid region into three elliptical island regions and a chaotic ``sea'', the dense jumble of particle locations seemingly without structure.  This common technique is also often called a \emph{Poincar\'{e} section}~\citep{Aref_frontiers_2017}.

The blue and red lines in Figure~\ref{fig:OrderWithinChaos} are structures termed respectively \emph{stable manifolds} and \emph{unstable manifolds}, which are associated with the central hyperbolic periodic point \citep{Ottino:1989aa}.  The stable manifold is the most repelling material curve, while the unstable manifold is the most attracting material curve.  While the stable and unstable manifolds are time inverses of each other, we focus on the red unstable manifold and note how much it resembles the deformed red blob, particularly the deformed red blob after 6 periods (Figure~\ref{fig:IllustrateChaospartd}).  This is no coincidence.  The unstable manifold is a fractal structure that fills the chaotic sea, and material starting near the central hyperbolic point is stretched along the unstable manifold and eventually converges to the unstable manifold.  The unstable manifolds of other hyperbolic points, if they exist, interweave together to form the chaotic seas.  The unstable manifold is the fundamental mixing pattern (or ``mixing template'', see \cite{Aref_frontiers_2017}).  Technically to generate chaotic advection requires {\em transverse intersection} of the unstable and stable manifolds, and manifold interaction is in fact the generic mechanism behind---and topological condition for---chaos in both steady and time-periodic flows.

\begin{figure}[tb]
    \centering
    \includegraphics[width=0.8\columnwidth]{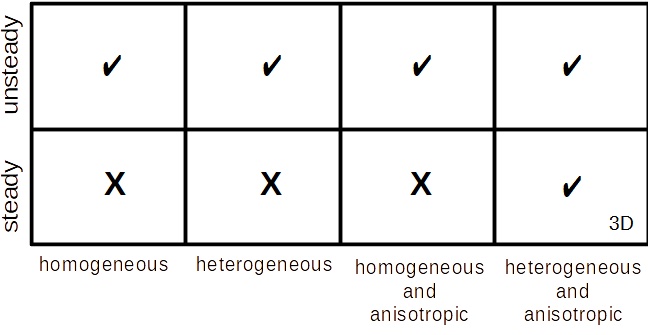}
    \caption{Diagram of when Darcy flows have the potential for chaos, depending on whether the hydraulic conductivity $\boldsymbol K$ is smooth or discontinuous, homogeneous, or heterogeneous, locally isotropic or anisotropic, and whether the flow is steady or unsteady in time.  Steady flows generally do not admit chaos, except when $\boldsymbol K$ is sufficiently complex, i.e. when the conductivity field is locally anisotropic or discontinuous.  Chaos is possible in all unsteady cases.} 
    \label{fig:Darcy_chaos_diagram}
\end{figure}

So far, we have given a very brief overview through one particular example of how Lagrangian critical regions, consisting of attracting, repelling, and hold-up regions along with the distinguished connections between critical regions, organize the Lagrangian transport of passive fluid particles. An immediate question is then where does this kind of transport occur in porous media flows?  The answer is wherever there are sufficient degrees of freedom (DOFs) in the fluid velocity field, be that as an upscaled continuum flow or in pore spaces, chaotic advection is the norm rather than the exception.  The sufficient DOFs for chaos in continuous systems is three, meaning that 3D steady or 2D unsteady flows can possibly exhibit chaotic advection~\citep{Speetjens_3D_2021}, whereas steady 2D flows are intrinsically non-chaotic. Thus we must reconsider if steady 2D flow approximations can adequately represent many porous media flows at both the pore and Darcy scales.

Consider first the Stokes flow in pore spaces discussed in section~\ref{sec:pores}.  Here the 3D pore spaces intrinsically have many hyperbolic points in the skin friction field of the solid matrix surface \citep{Lester:2013aa} that generically generate chaotic fluid trajectories.  Suggestive of this generic chaos is the similarity of the stretched and folded solute plume in the experiment of figure~\ref{fig:poremix}(c) with the stretched and folded filaments of Figure~\ref{fig:2dPeriodicExample}.  However, in 3D periodic or non-periodic flows not all of the organizing LCSs are as well characterised mathematically or experimentally as they are in 2D; there is a richer set of of Lagrangian structures and connections between them that one can investigate and use \citep{Haller:2015aa,Speetjens_3D_2021}.

These mechanisms also apply to evolution of the diffusive dye blob shown in Figure~\ref{fig:pushpullexperiment}, which is most clearly elucidated by first considering the push phase through the bead pack as a complex steady 3D flow. Unlike the time-periodic 2D flow considered in  Figure~\ref{fig:2dPeriodicExample}, there do not exist periodic points and lines in this flow, but there still exist stable and unstable manifolds (attracting and repelling LCS) that organise transport and impart exponential stretching of fluid elements. The diffusive dye shadows the unstable manifold of the push flow, and diffuses transverse to this structure in a filamentous manner whilst being stretched exponentially along the backbone of the unstable manifold. Eventually these filaments merge together via diffusion, leading to a well-mixed state. However, for the push-pull experiment, the flow is reversed before this state is reached. Due to reversal the stable and unstable manifolds are exchanged, and the dye blob now evolves along the stable manifolds of the push flow, leading to the highly striated structures shown in Figure~\ref{fig:pushpullexperiment} for different reversal times $\bar{t}$. While the original dye blob would be recovered upon flow reversal for non-diffusive dyes, for diffusive dyes, the irreversible nature of diffusion means that the resultant dye structure contains artifacts from the stretching history of the blob, and so provides a diffusive snapshot of the stable manifolds of the push flow.

Similarly, the Poincare sections shown in Figure~\ref{fig:varyC} clearly illustrate the LCS of the time-periodic poroelastic Darcy flow, where elliptic and hyperbolic points, elliptic islands and stable and unstable manifolds are shown. In this case, Lagrangian transport is trivial for small values of the medium compressibility $\mathcal{C}$ (Figure~\ref{fig:varyC}(a)), but with increasing compressibility a topological bifurcation occurs, leading to formation of elliptic and hyperbolic points (Figure~\ref{fig:varyC}(b,c)). Under a further increase of the medium compressibility (Figure~\ref{fig:varyC}(d)), the stable and unstable manifolds form a heteroclinic connection (not shown), and a chaotic sea forms around the elliptic island. These LCS govern the transport of both non-diffusive and diffusive species in this time-periodic poroelastic Darcy flow, and can lead to trapping of solutes, regions of accelerated mixing and transport that cannot be observed without elucidation of the LCS. In general, the tools and techniques associated with chaotic advection are necessary to understand solute transport in a wide range of porous media applications.

For Darcy type flows Figure~\ref{fig:Darcy_chaos_diagram} is a guide. Steady Darcy flows, even in 3D, are non-chaotic unless the conductivity field is sufficiently complex.  The constraint that causes this behaviour is discussed in section~\ref{sec:darcy}.  Unsteady Darcy flows in either 2D or 3D can always be made chaotic if one has control over the time-dependent flow forcing.  In natural Darcy flows, time-dependent forcing can unleash chaotic fluid trajectories, but as illustrated in Figure~\ref{fig:varyC}, the transition to chaotic behaviour is dependent upon the details of the system at hand.

To end this section we point out one more feature of Figure~\ref{fig:2dPeriodicExample}, our simple prototype chaotic flow:  the typical Lagrangian picture consists of both hold-up regions (fluid somewhat or completely isolated from the rest of the domain) and well-mixed regions (generated by the manifold interactions of the attracting and repelling regions).  Applications of porous media flows often require maximizing one or the other, i.e. isolating fluid or mixing it.  It is not always easy to find the particular flow forcing sequence that will maximize fluid isolation or mixing.  However, it can be done, and Figure~\ref{fig:Yap_confinement} shows an experiment designed to confine fluid in a Darcy flow, while Figure~\ref{fig:Cho_experiment} shows a field experiment designed to maximize mixing of emplaced fluid.  Developing general tools to take a porous media flow and quickly tailor it for a particular purpose is an area of ongoing research.

\section{Chaos in Pores}
\label{sec:pores}

\begin{figure}[tb]
\centering
\begin{tabular}{c}
\includegraphics[width=0.95\columnwidth]{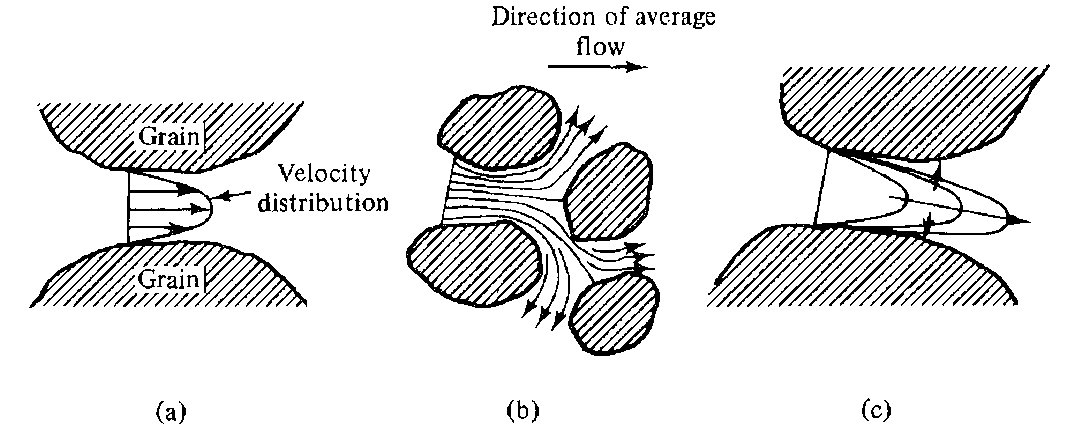} \\
\includegraphics[width=0.95\columnwidth]{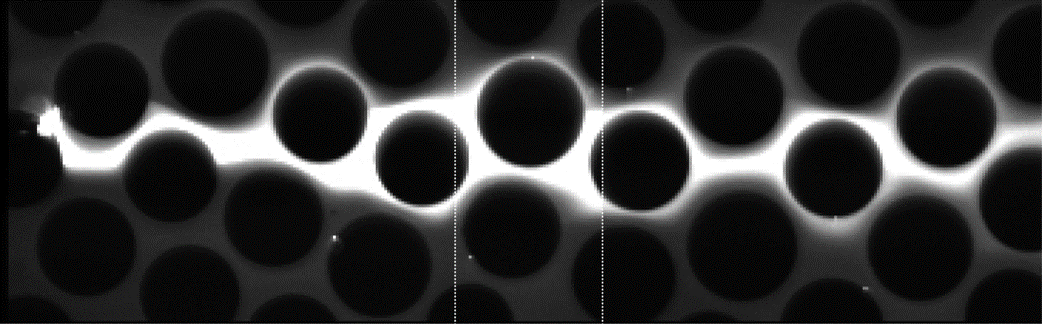}
\end{tabular}
\caption{The two-dimensional picture.  (top) Conventional depiction of hydrodynamic dispersion mechanisms operating at the pore scale (the graphic is reproduced from \cite{BearVerruijt:1987aa}): (a) mechanical dispersion via Poiseuille flow profiles at pore throats, (b) mechanical dispersion via flow separation at obstacles, (c) spreading due to molecular diffusion (Brownian motion).  (bottom) Experimental image of hydrodynamic dispersion in 2D random porous media (mean flow is from left to right) illustrating splitting and recombination of fluid streamlines. Adapted from \cite{Bruyne:2014aa}.}
\label{fig:2Dpicture}
\end{figure}

The classical picture of pore-scale fluid flow and solute dispersion follows a steady 2D paradigm as shown in the top part of Figure~\ref{fig:2Dpicture}, which is reproduced from \cite{BearVerruijt:1987aa}. This 2D conceptual image illustrates the main features that govern solute transport, including mechanical dispersion due to (a) strong flow velocity gradients between grains, (b) flow separation around grains and (c) molecular diffusion of the solute.  Although these solute transport mechanisms have been elaborated significantly from many contributions over the past few decades, it is sometimes overlooked that the 2D nature of this conceptual image itself imposes severe constraints on the admissible fluid dynamics, which in turn may limit our mental picture of solute transport. Although conceptual, these 2D topological constraints implicitly skew our understanding of the true 3D mechanisms which drive solute transport and sometimes lead us to neglect other important phenomena which may occur in 3D domains.


An important topological constraint that applies to all steady 2D flows is that streamlines cannot cross over each other, and so separating streamlines that pass either side of a grain must eventually recombine. Hence, the separation distance between neighbouring streamlines may only fluctuate as they move downstream; they cannot grow or shrink without bound. In mathematical terms, this topological constraint means that fluid elements in steady 2D flow may only grow at most algebraically with time~\citep{Dentz:2016aa} (due to e.g. shear between grains), which in turn places limitations on the rates of solute dispersion and mixing in such flows~\citep{LeBorgne:2015aa}. This constrained streamline behaviour in steady 2D flow also limits transverse dispersion as molecular diffusion is the only mechanism by which solutes can spread transversely to the mean flow direction. Conceptually, and in the absence of molecular diffusion, solute molecules simply follow initial streamlines and so cannot spread transversely, hence mechanical transverse dispersion is zero in 2D. Experimental observation of this 2D splitting-recombination phenomenon is shown in the bottom part of Figure~\ref{fig:2Dpicture}, where the solute plume in the 2D porous medium splits and recombines several times, yet the net transverse dispersion is limited and is governed by the solute molecular diffusivity. 

\begin{figure}[tbp]
\centering
\begin{tabular}{cc}
\multicolumn{2}{c}{\includegraphics[width=0.9\columnwidth]{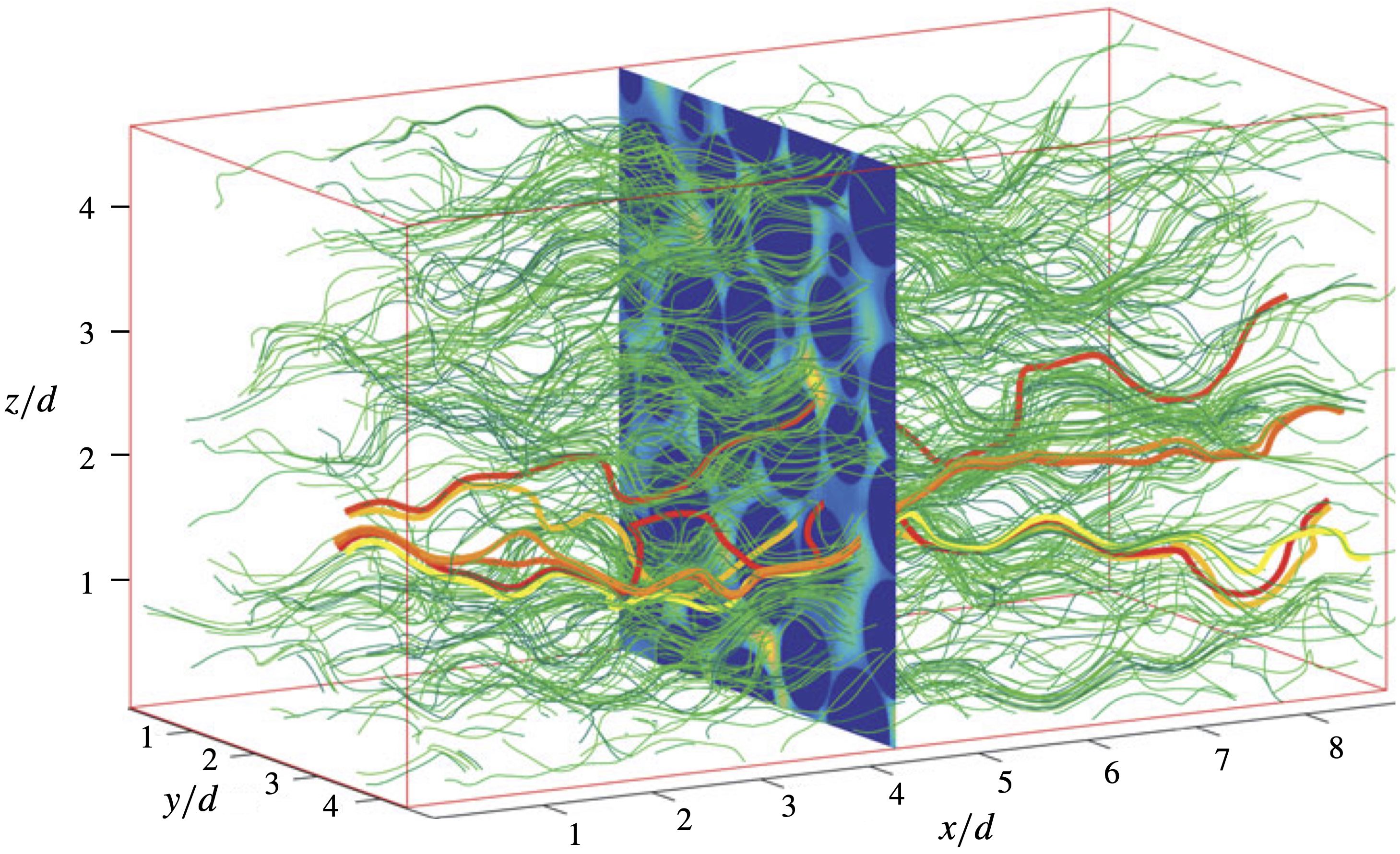}}\\
\includegraphics[width=0.38\columnwidth]{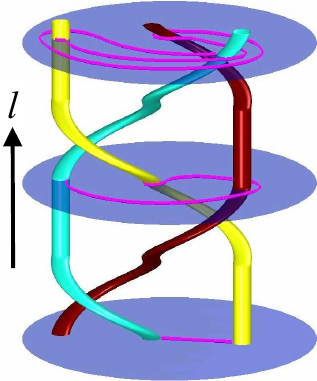} &
\includegraphics[width=0.43\columnwidth]{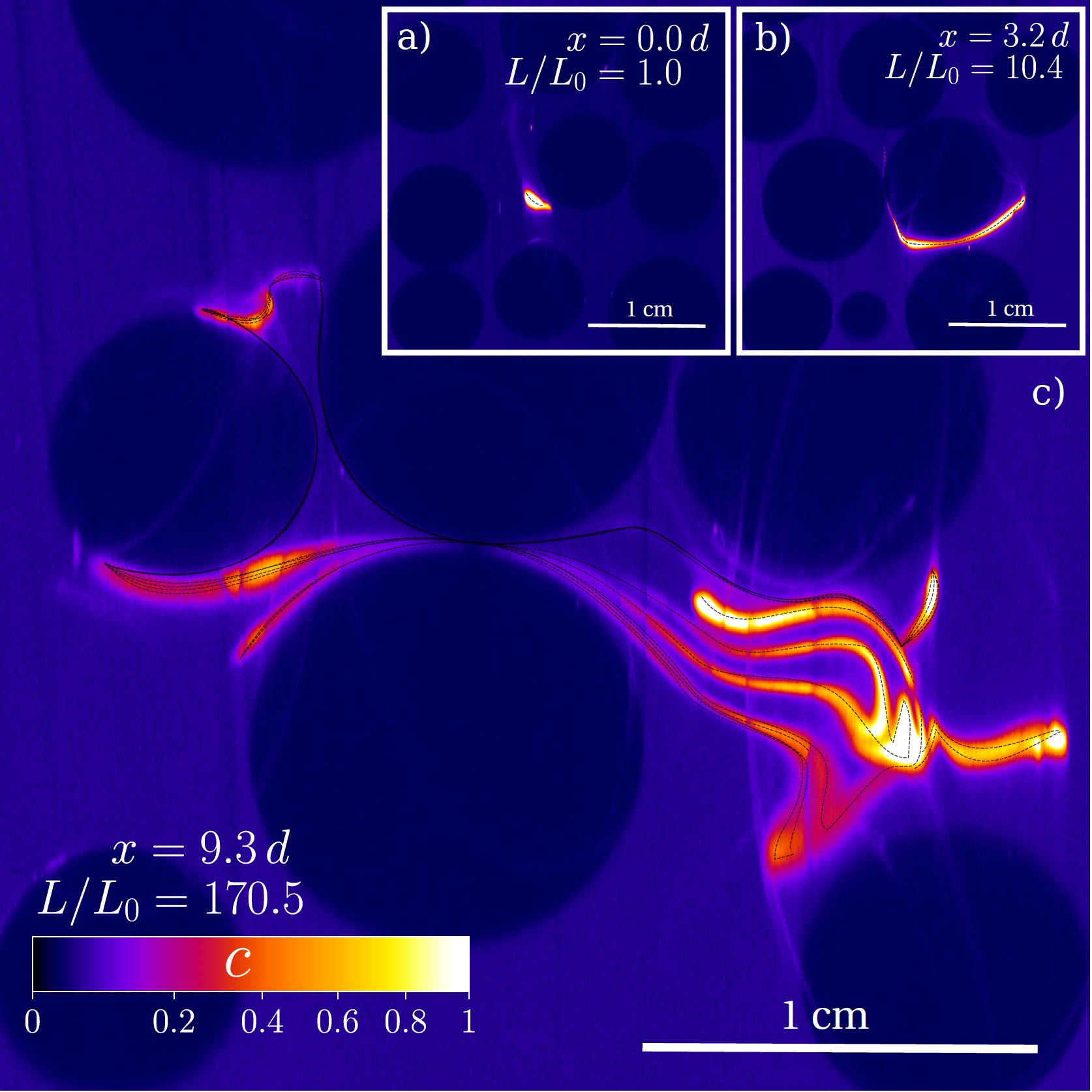}\\
(b) & (c)
\end{tabular}
\caption{(a) Numerically reconstructed streamlines from PIV experiments of steady 3D flow within a random glass bead pack illustrating complex braiding motions of tracer particles. Adapted from \cite{souzy:2020aa}. (b) Braiding motion of three streamlines (red, yellow, cyan) with downstream distance $l$ in a chaotic steady 3D flow. Due to the braiding motion of these streamlines, a material line (purple) connecting the yellow and cyan streamlines must grow exponentially with downstream distance. Adapted from \cite{Boyland:2000aa}. (c) Experimental images of a steady solute dye plume at various cross-sections (transverse to the mean flow direction) at distance $x$ downstream from the injection point. The cross-sectional length $L$ of the plume grows exponentially with $x$ due to stretching and folding of fluid elements. Adapted from \cite{Heyman:2020aa}}
\label{fig:poremix}
\end{figure}

Although it is tempting to conceptualise steady flow through 3D porous architectures as an extrusion of the dynamics in Figure~\ref{fig:2Dpicture}, these topological constraints do not apply to 3D steady flows as the additional degree of freedom---the third spatial dimension---removes the restriction that a fluid streamline is bounded by its neighbours.  The result is that the streamline may wander freely throughout the fluid domain, as shown in Figure~\ref{fig:poremix}(a): in steady 3D flows neighbouring streamlines may diverge without bound as they are advected downstream, leading to wholly new transport mechanisms (such as chaotic advection) which are not readily apparent in conventional dispersion models. This freedom also admits braiding of streamlines, depicted in Figure~\ref{fig:poremix}(a), that is associated with exponential growth of transverse material lines.

The origin of chaotic advection in 3D pore-scale flows can be traced back to stagnation points that arise on grains or at pore boundaries in all porous media~\citep{Lester:2013ac}, and fluid deformation local to these points generates persistent exponential stretching of fluid elements in the bulk. Conversely, in steady 2D flows, such local fluid stretching at stagnation points is cancelled out at downstream reattachment points. The persistent exponential stretching in 3D leads to the highly striated and filamentous solute distributions transverse to the mean flow direction shown in Figure~\ref{fig:poremix}(c). Flow separation and persistent exponential stretching of material elements is wholly responsible for the complex steady solute plume structures observed in 3D pore-scale flows.  As these dynamics are inadmissible in steady 2D flow, they can be overlooked in conventional modelling approaches.

It is important to note that exponential fluid stretching qualitatively augments solute mixing, as these mixing dynamics are \emph{singular} in the presence of exponential fluid stretching~\citep{Cerbelli:2017aa} in that the rate of mixing in the limit of vanishing diffusivity is not the same as that for purely advective flow. Conversely, sub-exponential stretching (which is characteristic of regular fluid flow) exhibits non-singular mixing rates which limit to zero with vanishing molecular diffusivity. Hence, the impact of chaotic advection on solute mixing is profound.

The impact of pore-scale chaotic advection extends beyond solute mixing, and also acts to augment longitudinal dispersion~\citep{Lester:2014ab,Lester_anomalous_2014}, chemical and biological activity~\citep{Tel2005} and transport of colloids and micro-organisms~\citep{Haller2011aa}. As the P\'{e}clet number $Pe$ (which characterises the relative timescales of advection to diffusion) is typically small in pore-scale flows, it is commonly assumed that diffusion is rapid compared to advective transport and so chaotic advection has minimal impact. However, \cite{Heyman:2020aa} showed that exponential fluid stretching results in incomplete solute mixing at the pore scale even for small P\'{e}clet numbers ($Pe>5$): chaotic advection acts to augment diffusive mixing and transport across a wide range of pore-scale systems.

\begin{figure}[tbp]
\centering
\begin{subfigure}[b]{0.45\textwidth}
\includegraphics[width=\textwidth,height=\textwidth]{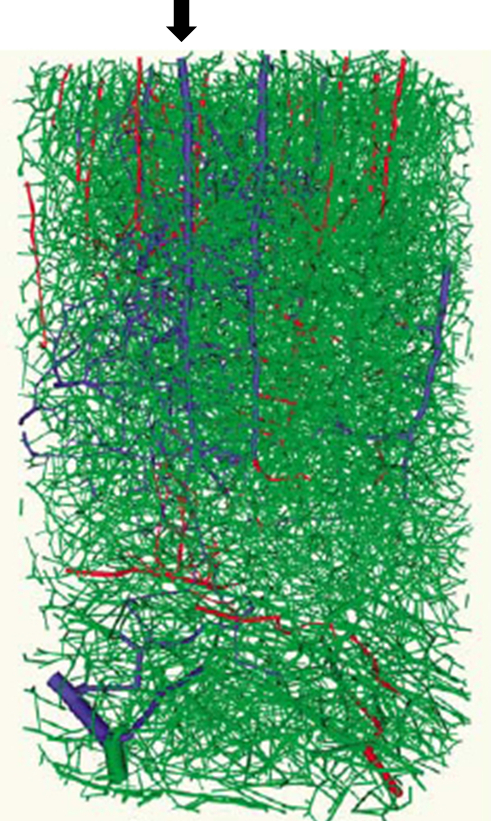}
\label{TransportLivingTissue2a}
\caption{}
\end{subfigure}
\begin{subfigure}[b]{0.45\textwidth}
\includegraphics[width=\textwidth,height=\textwidth]{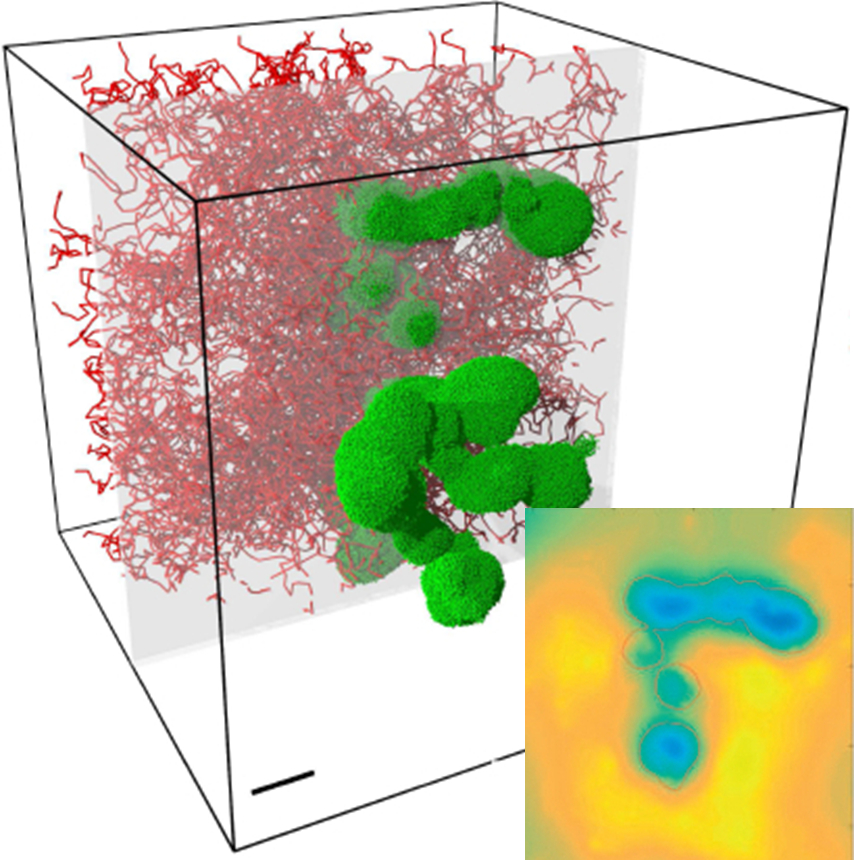}
\label{TransportLivingTissue2b}
\caption{}
\end{subfigure}
\caption{Micro-circulation in living tissue as physiological instance of transport in porous networks: (a) impact of vascular occlusion (arrow) on global through flow (from bottom) in somatosensory cortex in rat brain visualised by simulated flow rate (red/blue: relatively larger/lower flow rate; green: unaffected region) (adapted from \citep{Reichold2009}); (b)
impact of tumor cells (green) on oxygen distribution via vascular network (red) visualised by simulated oxygen levels (yellow/blue: healthy/hypoxic)
in grey cross section (inset) (adapted from \citep{Ghaffarizadeh2016}).}
\label{TransportLivingTissue2}
\end{figure}

\begin{figure}[tbp]
\centering
\begin{subfigure}[b]{0.45\textwidth}
\includegraphics[width=\textwidth,height=\textwidth]{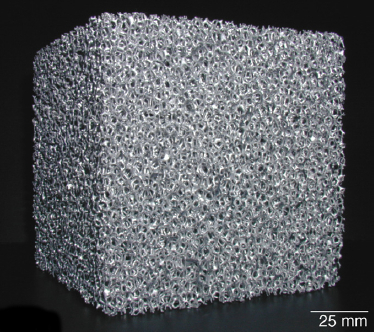}
\label{MetalFoamHX_a}
\caption{}
\end{subfigure}
\begin{subfigure}[b]{0.45\textwidth}
\includegraphics[width=\textwidth,height=\textwidth]{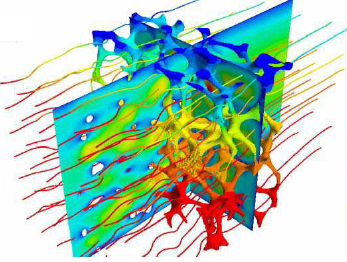}
\label{MetalFoamHX_b}
\caption{}
\end{subfigure}
\caption{Transport in engineered porous media illustrated by a compact heat exchanger based on metal foam: (a) typical metal foam (Fe-Cr-Al-Y alloy) for heat-transfer purposes (reproduced from \cite{Zhao2012}); (b) simulated cooling of hot fluid flow entering (from the left) metal foam with cold top and adiabatic bottom visualised by temperature of foam surface and streamlines (blue/red: min/max) and velocity magnitude (blue/red: zero/max) in planar cross-sections (adapted from \cite{Hugo:2011aa}).}
\label{MetalFoamHX}
\end{figure}

\cite{Heyman:2021aa} has recently developed a novel method to characterise pore-scale chaotic mixing in opaque porous media based on the push-pull flow shown in Figure~\ref{fig:pushpullexperiment}. This method has been used to recover the extent of chaotic mixing in model systems such as glass beads, as well as to estimate the extent of chaotic mixing in natural systems such as randomly packed gravel. This technique has the potential to characterise chaotic mixing across a broad range of natural porous media, ranging from granular matter to biological pore networks. Biological tissues may be regarded as porous networks that can host chaotic mixing, with applications in a range of contexts including living brain tissue~\citep{Reichold2009} (Figure~\ref{TransportLivingTissue2}a), vascular networks~\citep{Ghaffarizadeh2016} (Figure~\ref{TransportLivingTissue2}b) and lung alveoli~\citep{Tsuda:2002,Tsuda:2011}.  Conversely, \cite{Turuban:2018aa,Turuban:2019aa} show that highly ordered and engineered porous materials such as crystalline lattices of uniform spheres can exhibit a broad range of pore-scale mixing dynamics ranging from non-chaotic to chaotic mixing that is stronger than that exhibited by any known random system. These mixing dynamics strongly depend upon the orientation of the crystalline axes with respect to the mean flow direction.
\cite{Zhao2012} have used disordered engineered porous metal foams (Figure~\ref{MetalFoamHX}) to develop compact heat exchangers that accelerate heat transfer by generating chaotic mixing in the pore space. These advances illustrate the potential for chaotic mixing principles to assist the development of novel porous materials exhibiting tuneable mixing and transport properties, with myriad industrial and medical applications.

\section{Chaos in Darcy Flows}
\label{sec:darcy}

\subsection{Steady Darcy Flows}
\label{subsec:steady_darcy}

At scales much larger than the pore scale, it is neither practical nor desirable to resolve the detailed pore-scale flows described in the previous section. Under the assumption that the characteristic length scale $L$ of material properties or pressure fluctuations at the Darcy scale is much larger than the characteristic pore scale $\ell$, fluid flow in porous media is described by Darcy's law, which arises from a formal upscaling of the Stokes equation at the pore scale
\begin{align}
    \mu\nabla^2\mathbf{v}-\nabla p=0,\,\,\nabla\cdot\mathbf{v}=0,\,\,\,\,\mathbf{x}\in\Omega_f,\label{eqn:Stokes}
\end{align}
coupled with the no-slip boundary condition
\begin{align}
    \mathbf{v}=\mathbf{0},\,\,\,\,\mathbf{x}\in\partial\Omega_{fs},\label{eqn:noslip}
\end{align}
where $\mu$ is the fluid viscosity, $\mathbf{v}$ and $p$ respectively are the pore-scale velocity and pressure. The fluid-filled pore space is denoted $\Omega_f$, and the pore boundary by $\partial\Omega_{fs}$. 

Upscaling these equations via either volume averaging, method of moments or multiscale expansions recovers Darcy's law, an empirical law that links the macroscopic (averaged) Darcy flux $\mathbf{q}\equiv\langle\mathbf{v}\rangle$ to the macroscopic pressure field $P\equiv\langle p\rangle$ as
\begin{equation}
    \mathbf{q}=-\frac{\mathbf{K}}{\mu}\cdot\nabla P,\quad \nabla\cdot\mathbf{q}=0,\,\,\,\,\mathbf{x}\in\Omega.\label{eqn:Darcy}
\end{equation}
$\mathbf{K}$ is the permeability tensor (and $\mathbf{K}/\mu$ is known as the hydraulic conductivity tensor) and $\Omega=\Omega_f\cup\Omega_s$ denotes the \emph{porous medium} 
that is comprised of both the fluid-filled pore-space $\Omega_f$ and the solid matrix $\Omega_s$.  This upscaling process effectively removes the pore boundary $\partial\Omega_{fs}$ and the no-slip condition (\ref{eqn:noslip}).  Now the permeability tensor $\mathbf{K}$ accounts for the flow resistance due to no-slip. In this way, $\mathbf{K}$ may be considered a closure model for the Darcy equation (\ref{eqn:Darcy}) as a result of upscaling Stokes flow at the pore scale. For many classes of porous media, the permeability is isotropic and may be represented by the scalar value as $\mathbf{K}=k\mathbf{I}$ where $\mathbf{I}$ is the identity tensor, and Darcy's law simplifies to
\begin{equation}
    \mathbf{q}=-\frac{k}{\mu}\nabla P,\quad \nabla\cdot\mathbf{q}=0,\mathbf{x}\in\Omega.\label{eqn:Darcy_scalar}
\end{equation}
As the pore boundary is key to the generation of pore-scale chaos, omission of this boundary in the upscaling process also omits the dynamics associated with pore-scale chaos. Thus the impacts of pore-scale mixing on solute dilution, dispersion and transport must be incorporated via appropriate upscaled models. For homogeneous porous media where the isotropic permeability $k$ is constant, steady 3D Darcy flows are irrotational (as the vorticity is identically zero: $\boldsymbol\omega\equiv\nabla\times\mathbf{q}=-k/\mu(\nabla\times\nabla P)=0$), and so must also be non-chaotic as there is no mechanism for the creation of chaos-inducing structures known as homoclinic and heteroclinic transverse connections (discussed earlier). Similarly, even for strongly heterogeneous porous media with smooth, isotopic conductivity, i.e. $k=k(\mathbf{x})$, steady Darcy flow is non-chaotic as the helicity density defined~\citep{moffatt_1969} as the dot product of Darcy flux and vorticity is identically zero:
\begin{equation}
    h(\mathbf{x})\equiv\mathbf{q}\cdot\boldsymbol\omega=k/\mu^2\nabla P\cdot(\nabla P\times\nabla k)=0.\label{eqn:helicity}
\end{equation} The helicity density is a measure of the topological complexity of a flow, and it can be shown~\citep{Arnold:1965} that all steady helicity-free flows are non-chaotic.

\begin{figure}[tbp]
\centering
\begin{tabular}{cc}
\includegraphics[width=0.45\columnwidth]{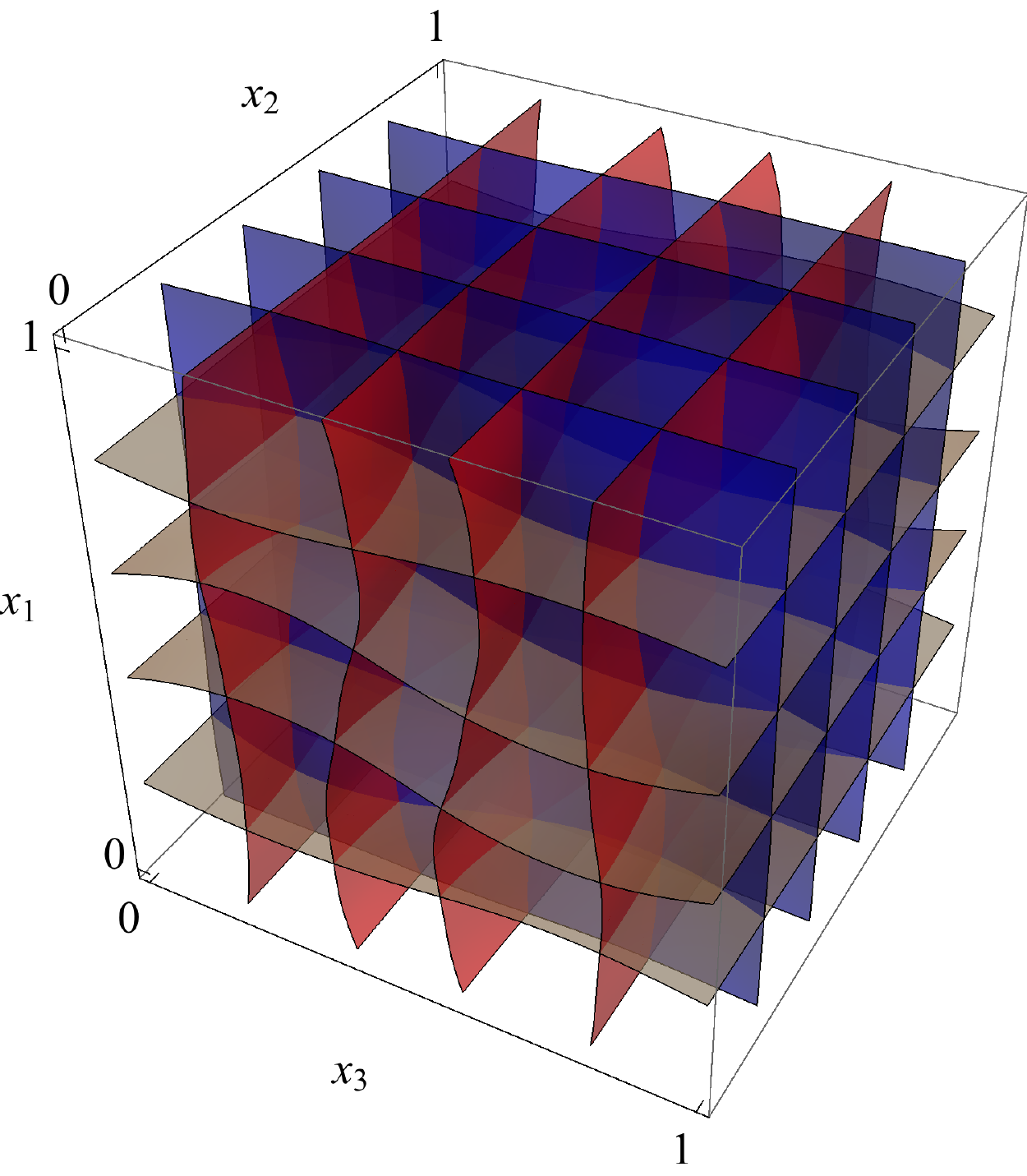} &
\includegraphics[width=0.45\columnwidth]{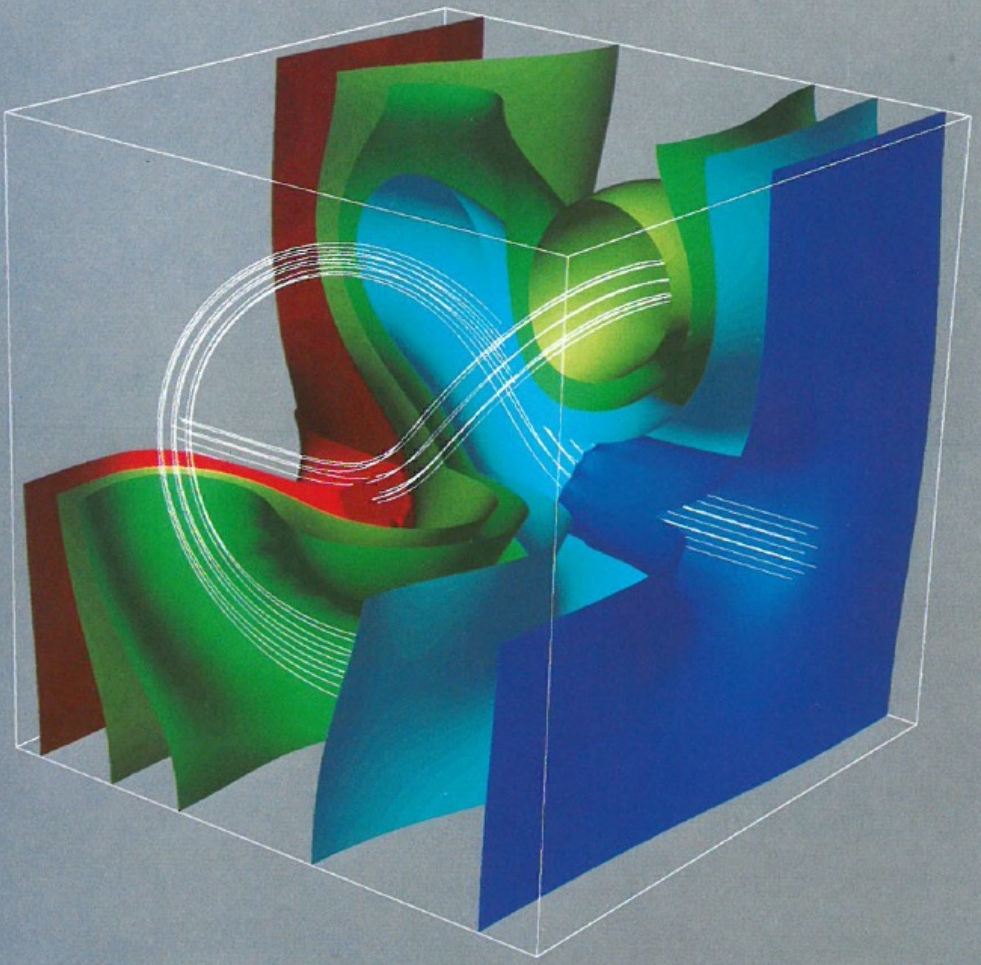} \\
(a) & (b)
\end{tabular}
\caption{(a) Complex of pressure $P$ (grey) and streamfunction $\psi_1$ (red), $\psi_2$ (blue) isosurfaces for a steady 3D Darcy flow in a porous medium with heterogeneous, smooth and isotropic hydraulic conductivity. Streamlines arise as the intersection of 2D topologically planar streamsurfaces, and so cannot exhibit braiding motions characteristic of chaotic flow. Adapted from \cite{Lester:2021aa}. (b) Knotted fluid particle trajectories (white lines) and isoconductivity surfaces (coloured surfaces) in a steady anisotropic 3D Darcy flow. Adapted from \cite{Cole:1990aa}.}
\label{fig:Darcy}
\end{figure}

In a series of pioneering studies~\cite{Sposito:1997aa,Sposito:2001aa,Sposito:2006aa} argued that the helicity-free condition (\ref{eqn:helicity}) gives rise to \emph{Lamb surfaces}~\citep{Lamb:1932aa} that are spanned by streamlines and vorticity lines of the flow, and provide strong topological constraints on streamline motion. \cite{Lester:2019aa} showed that Lamb surfaces only exist for trivial isotropic 3D Darcy flows; however, it has since been established \citep{Lester:2021aa} that the helicity-free condition admits a pair of coherent streamfunctions $\psi_1(\mathbf{x})$, $\psi_2(\mathbf{x})$ ~\citep{Zijl:1986aa}. These streamsurfaces $\psi_i(\mathbf{x})=\text{constant}$ are ``foliated'' (stacked together) throughout the flow as non-intersecting 2D lamellae.

The intersection of these 2D streamsurfaces form the 1D streamlines of the flow, which are topologically equivalent to straight lines. As shown in Figure~\ref{fig:Darcy}a, even if the conductivity field is strongly heterogeneous, these 1D streamlines must be confined to these topologically planar 2D streamsurfaces and so are topologically equivalent to parallel straight lines. Thus, 1D streamlines in steady isotropic flow cannot undergo the braiding motions (shown in Figure~\ref{fig:poremix}b) that are characteristic of chaotic mixing. Thus, no matter how heterogeneous, steady isotropic 3D Darcy flows have trivial flow topology akin to that of steady 2D flow. \citet{Lester:2018aa,Lester:2021ab} show that this simplified flow topology leads to algebraic rates of fluid deformation, solute mixing and dispersion in isotropic 3D Darcy flow, and so these phenomena are qualitatively similar to those of steady 2D Darcy flow.

The topological simplicity of steady isotropic Darcy flows can be broken in a number of ways, leading to the possibility of more complex flow topology and chaotic mixing. First, if the Darcy flow is unsteady, the trajectories of fluid tracer particles are no longer confined to the streamlines of the steady flow and so the flow topology can be much richer and possibly chaotic~\citep{weeks:1998aa}. Second, if the permeability field is discontinuous, the kinematic constraints associated with helicity-free flows not longer apply. This is the case for Darcy flow in isotropic heterogeneous porous media with spheroidal inclusions \citep{Jankovic:2003aa,Jankovic:2009aa,DiDato:2016aa,DiDato:2016ab}, as the conductivity field is discontinuous at the inclusion boundaries, and so the vorticity and helicity is undefined. This relaxation of the helicity-free constraint  leads to observations of non-trivial streamline braiding similar to that shown in Figure~\ref{fig:poremix}b. Third, if the permeability tensor is locally anisotropic and heterogeneous, i.e. $\mathbf{K}=\mathbf{K}(\mathbf{x})$, then the helicity density is non-zero in general:
\begin{equation}
    h(\mathbf{x})\equiv\mathbf{q}\cdot\boldsymbol\omega=(\mathbf{K}\cdot\nabla P)\cdot(\nabla\times\mathbf{K})\cdot\nabla P\neq 0.\label{eqn:helicity_tensor}
\end{equation}
Breaking the zero helicity topological constraint means that the coherent streamfunctions $\psi_1$, $\psi_2$ no longer exist and streamlines are no longer confined to topologically planar 2D streamsurfaces but rather are free to wander throughout the flow domain. This topological freedom opens the possibility for the braiding of streamlines (as shown in Figure~\ref{fig:poremix}b), exponential stretching of material elements and chaotic motion. This potential topological complexity is reflected in Figure~\ref{fig:Darcy}b, which shows clear evidence of knotted and braided streamlines that are characteristic of chaotic mixing. Several studies~\citep{Cirpka:2015aa,Chiogna:2014aa,Chiogna:2015aa,Ye:2015aa} have observed complex mixing dynamics and streamline structure in anisotropic Darcy flows, however the signatures of chaotic mixing (such as positive Lyapunov exponent or non-trivial flow topology) are yet to be identified in these flows. At the Darcy scale, as chaotic mixing is anticipated to have a profound impact upon solute mixing, transport and dispersion, uncovering these dynamics and their link to the structure of the anisotropic permeability tensor is currently an important open problem in groundwater hydrology.

At regional scales much larger than the Darcy scale, it is often not computationally feasible to completely resolve the heterogeneous permeability field, and upscaling techniques such as block or effective conductivity are employed~\citep{SanchezVila:2005aa}. In many cases the locally isotropic (scalar) permeability field $k(\mathbf{x})$ at the Darcy scale is statistically anisotropic (in that the correlation structure is orientation-dependent due to e.g. geological layering etc), and in these cases, conventional upscaling techniques~\citep{SanchezVila:2005aa} result in an upscaled permeability field $\mathbf{K}^\star$ which is tensorial. \cite{Cirpka:2015aa} and \cite{Chiogna:2014aa} have shown that although the fully resolved Darcy-scale flow is helicity-free and has trivial mixing dynamics, the upscaled regional flows can have non-zero helicity and complex streamline motion. In contrast to upscaling from the pore scale to the Darcy scale, where pore scale chaotic dynamics are omitted via the upscaling process, these results suggest that upscaling from the Darcy scale to regional scales may introduce chaotic dynamics that are not apparent at the Darcy scale. As upscaling is essentially a coarse-graining process that results in a loss of information, it is argued that these chaotic dynamics are a non-physical artefact of the employed upscaling methods employed. This is a research area that requires further investigation and may call for \emph{topologically consistent} upscaling methods that preserve the Lagrangian topology of the fully resolved Darcy-scale flow.

\subsection{Transient Darcy Flows}
\label{subsec:transient}

\begin{figure}[tbp]
    \centering
\includegraphics[width=1.0\textwidth]{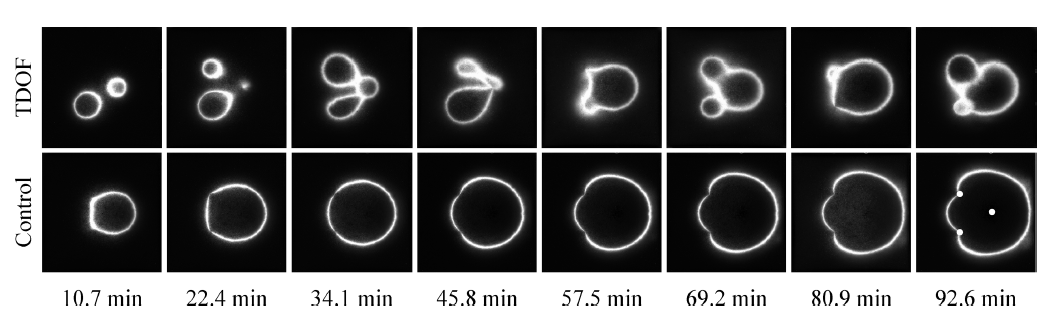}    
    \caption{Reactive mixing under (top) transient and (bottom) steady Darcy flows. A colorimetric reaction occurs at the interface of a clear tiron (Ti) solution that is injected into the flow domain (via the three wells are indicated as white dots on the last image) to displace an existing clear molybdate (Mo) solution to form the stable coloured product MoTi$_2^{4+}$. These images clearly show that time-dependent pumping can rapidly accelerate scalar mixing and reaction over the steady case. From \protect\cite{Zhang:2009aa}.}
    \label{fig:Bagtzoglou_reaction}
\end{figure}

Transient Darcy flows can not only accelerate transport and dispersion~\citep{Dentz:2005aa,Cirpka:2003aa}, but in certain cases can also admit a range of qualitative transport behaviours that may not otherwise occur in their steady counterparts~\citep{weeks:1998aa}. These transport dynamics are critical to the transport and fate of solutes and colloids in both natural and engineered systems, from seasonal and tidal forcing in coastal aquifers, through to transport through biological tissues and porous engineered structures. Due to the unsteady nature of these flows, it is difficult to detect and understand the transport structures in the Eulerian frame, whereas the Lagrangian frame facilitates detection and classification of the kinematic features that can control complex fluid mixing, dispersion, reaction, segregation and discharge phenomena at the Darcy scale. Together, these kinds of transport phenomena describe complex Lagrangian structures which engender potentially profound impacts on solute migration and reaction \citep{Toroczkai:1998aa,Tel2005,Valocchi:2019aa}. Building on early conceptual work by \citet{Jones:1994aa}, \citet{Ottino:1989aa}, \citet{Sposito:2001aa,Sposito:1997aa} and others \citep{Piscopo:2013aa,Lester:2009ab,Lester:2010ab,Metcalfe:2010aa,Trefry:2012aa,Mays:2012,Lester:2016ab}, groundwater researchers are now able to predict, observe and engineer complex Lagrangian structures in saturated porous media at the laboratory \citep{Zhang:2009aa,Metcalfe:2010aa,Bagtzoglou:2007aa} and field \citep{Cho:2019aa} scales, with quantified benefits for mixing and reactivity enhancement.

\begin{figure}[tbp]
    \centering
\begin{tabular}{cc}
\includegraphics[width=0.45\textwidth]{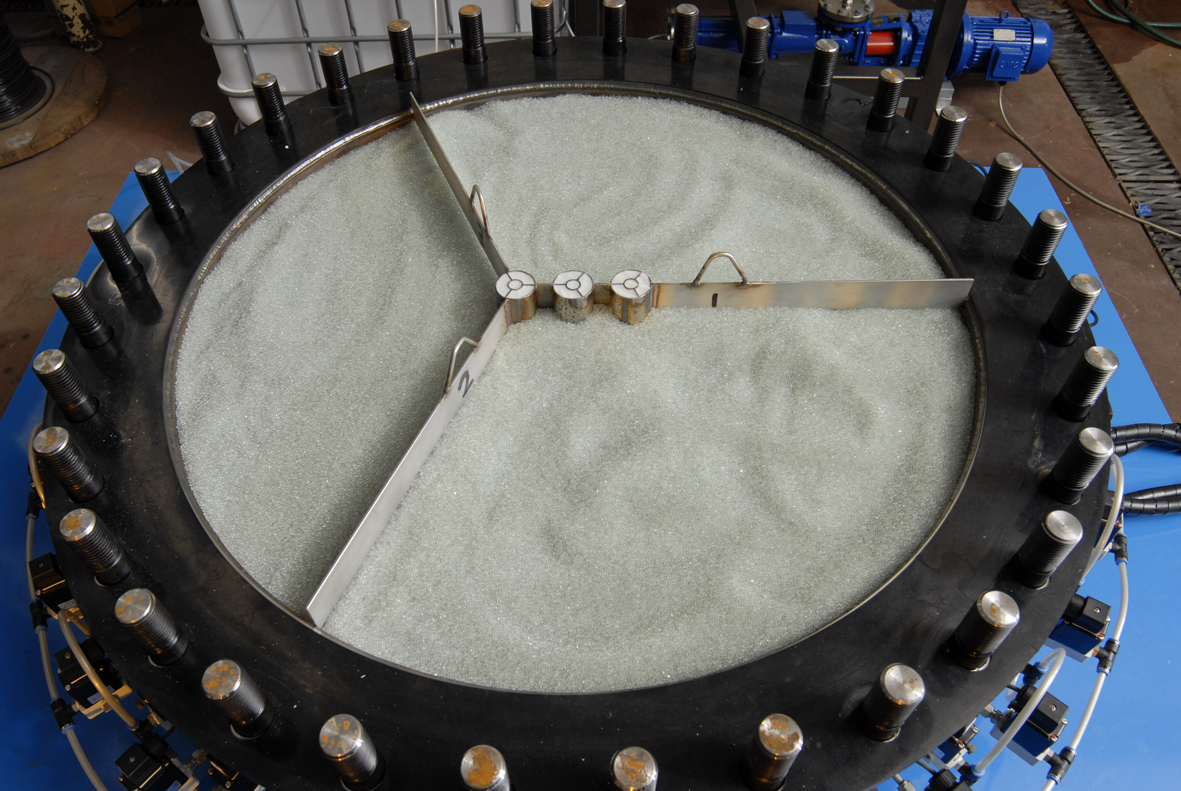}  & \includegraphics[width=0.45\textwidth]{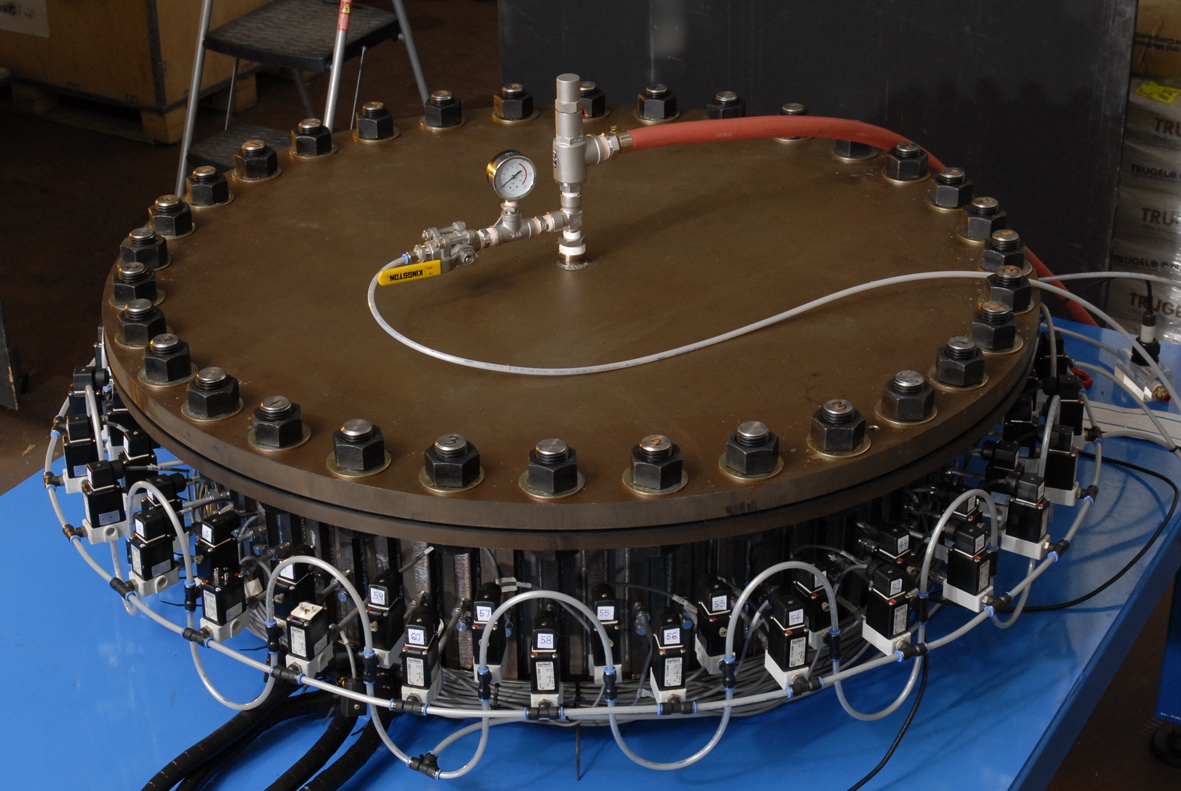} \\
(a) & (b) \\
\multicolumn{2}{c}{\includegraphics[width=\textwidth]{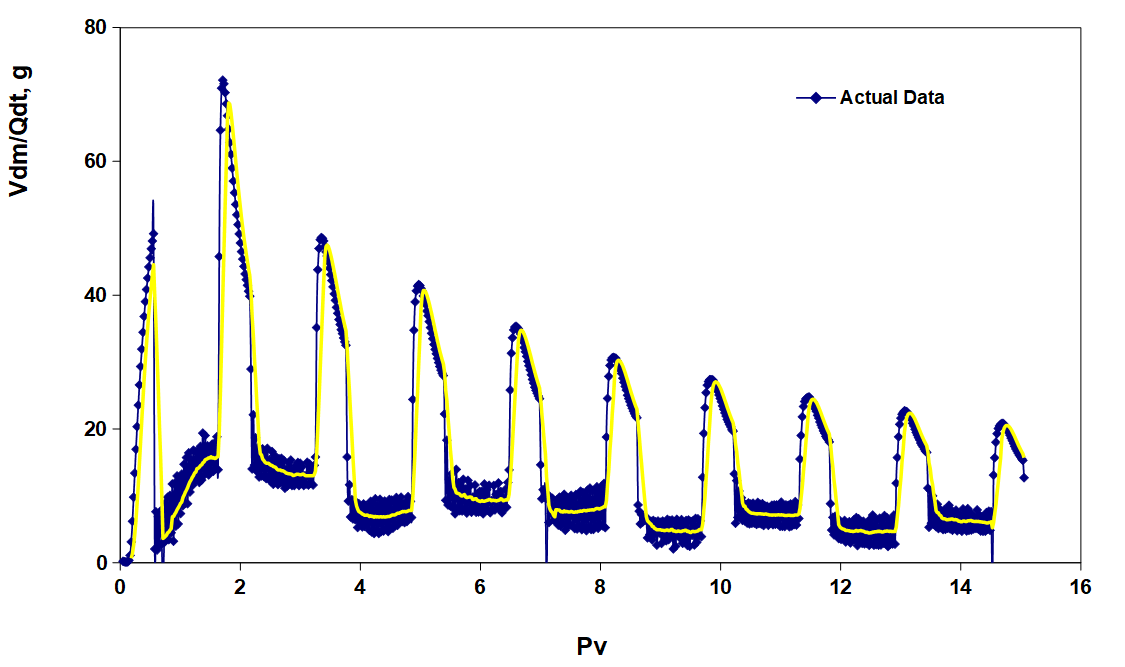}} \\
\multicolumn{2}{c}{(c)}
\end{tabular}
    \caption{Mesoscale solute confinement experiment. (a) A 1.5m diameter cylindrical cavity is filled with glass beads, and three pillars of salt (centre) are surrounded by smaller glass beads. A series of fluid injection and extraction ports around the perimeter (b) facilitate generation of different rotated potential mixing (RPM) flows. (c) Repeated switching between confinement and mixing protocols of the RPM flow correlate well with regions of high and low salt concentration exiting the device.  $P_v$, pore volumes of fluid moved through the domain, is a measure of elapsed time. Adapted from \citet{Trefry:2012ac}.}
    \label{fig:Yap_confinement}
\end{figure}

\begin{figure}[tbp]
    \centering
\begin{tabular}{cc}
\includegraphics[width=0.3\textwidth]{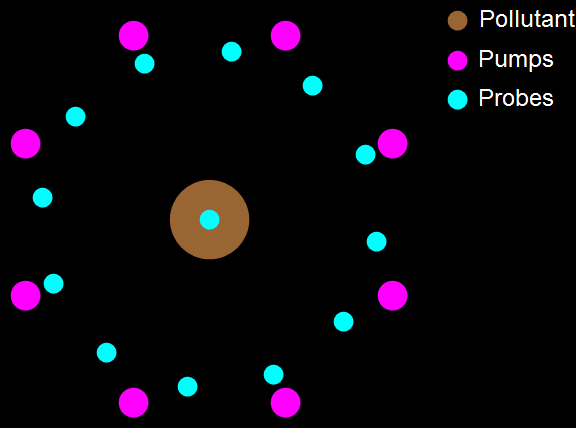}
 & \includegraphics[width=0.63\textwidth]{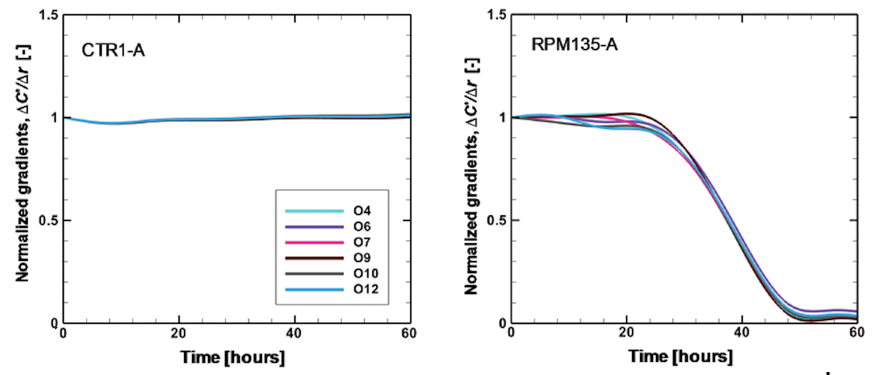}\\
\end{tabular}   
    \caption{Subsurface chaotic mixing field experiment at the Borden Research Aquifer. (left) Schematic of the experiment, with a centrally emplaced ``pollutant'' of salt water, pumps to generate time-dependent injection/extraction well pairs, and probes for electrical resistance tomography measurement of the salt concentration field and to measure the radial concentration gradient as a function of angle. Normalized radial concentration gradients as a function of time for (middle) a control experiment without mixing and (right) an experiment with a particular RPM mixing protocol. Adapted from  \citep{Cho:2019aa}}.
    \label{fig:Cho_experiment}
\end{figure}

\citet{Zhang:2009aa} clearly demonstrated (Figure~\ref{fig:Bagtzoglou_reaction}) that a significant increase in reaction rate and yield can be achieved when engineered chaotic mixing flows are introduced in laboratory-scale experiments. Taking this idea further, \cite{Metcalfe:2012ab} and \cite{Trefry:2012ac} used a novel pumping arrangement based upon the reoriented potential mixing (RPM) flow~\citep{Metcalfe:2010aa,Metcalfe:2010ab,Lester:2009ab} to demonstrate control of solute transport at the mesoscale (Figure~\ref{fig:Yap_confinement}), alternating between confinement of a solute plume and accelerated mixing and release. \cite{Cho:2019aa} then demonstrated that chaotic mixing could be also used at the field scale to accelerate solute transport and mixing (Figure~\ref{fig:Cho_experiment}), with obvious applications in pollutant remediation and resource extraction. 

Complex Lagrangian structures have also been predicted to occur in natural groundwater environments, i.e. Darcian systems where engineered pumping activity is absent. Using a conventional Darcian confined flow model in two dimensions (vertically averaged),
\cite{Trefry:2019aa} showed computationally that complex Lagrangian structures arose where aquifer heterogeneity and compressibility were combined with sinusoidal boundary forcing. This analysis was extended by \cite{Wu:2019aa} to consider wide ranges of relevant parameter values, highlighting that the spectra of natural ocean tides around the globe provides significant potential to induce Lagrangian complexity in coastal aquifer flows; other important hydraulic parameters are the aquifer compressibility (as depicted in Figure~\ref{fig:varyC}), diffusivity and heterogeneity. These complex mixing dynamics have also been shown~\citep{Trefry:2020aa} to extend to more realistic (and complex) scenarios where the boundary forcing is comprised of several superposed periodic modes. 
Since the minimal ingredients to porous media LCS---heterogeneity, compressibility and time-dependent forcing---arise in other geophysical, industrial and biological systems, the above rich set of transport structures that determine fluid residence times, solute segregation, mixing and reaction can also arise in a wide range of other porous media applications.

\section{Applications and Consequences}

In previous sections we have seen how dynamical systems approaches allow us to detect, classify and quantify the various kinematic structures inherent in a wide array of physical/biological systems and across dimensionalities. These outcomes are not only conceptual --- they also provide us with information that can be applied to old and new problems for benefit. Here we discuss in greater depth two examples where the study of Lagrangian kinematics can led to advances in understanding and quantification of fundamental transport and mixing processes, and optimisation of engineering design and performance for porous systems.

\subsection{Pore Scale --- Fluid Deformation and Solute Mixing}

Recent advances in imaging of pore-scale architecture, flow and solute transport have provided rich and detailed data sets regarding transport processes at the pore-scale. For example, several studies \citep{Arthur:2009aa,Yiotis:2021aa,Holzner:2015aa,souzy:2020aa,Linzhu:2022aa,Ahkami:2019aa} have employed 3D pore-scale particle image velocimetry (PIV) in optically transparent media to provide fully resolved measurements of the steady 3D pore-scale flow field. Motivated by observations of spatial Markovianity in heterogeneous porous media~\citep{Le-Borgne:2008aa,Le-Borgne:2008ab}, these data sets have subsequently been used to develop stochastic (CTRW) models of velocity evolution along streamlines, which governs processes such as longitudinal dispersion and contaminant breakthrough~\citep{Fouxon:2016aa,Bijeljic:2006aa,Le-Borgne:2011aa,Puyguiraud:2021aa}. In addition to the evolution of streamwise velocity, the deformation of fluid elements at the pore-scale govern a wide array of fluid-borne phenomena ranging from solute mixing and dispersion to chemical reactions and colloid transport. Given the deluge of PIV data regarding pore-scale flow, it can be difficult to know how to visualise these Lagrangian kinematics, let alone develop faithful models of fluid deformation. Such deformation is tensorial in nature (i.e. it cannot be quantified by a single scalar) in that fluid stretching, shearing and folding both parallel and transverse to streamlines governs different processes such as longitudinal and transverse mixing and dispersion~\citep{Dentz:2016aa}.

As fluid deformation is governed solely by Lagrangian kinematics, dynamical system techniques provide a means to visualise, classify and quantify fluid deformation driven by the organisation (specifically braiding and spreading) of streamlines in steady pore-scale flow~\citep{Lester:2016ab}.
Despite the random nature of these flows, it is necessary to develop stochastic models of fluid deformation that honour the Lagrangian kinematics and associated constraints. In this regard, the dynamical systems approach to the evolution of the fluid deformation gradient tensor provides a framework~\citep{Lester:2018aa,Lester:2021ab} and measurement techniques for the quantification of the transverse and longitudinal shear rates and Lyapunov spectra. Dynamical systems theory proves that for steady 3D flow, the Lyapunov exponents must sum to zero and have one zero component, hence the other two exponents must be of equal magnitude and opposite sign. Hence the stretching dynamics are fully characterised by the single positive Lyapunov exponent. On the basis of spatial Markovianity, these dynamical systems tools have been employed to develop stochastic models of the full fluid deformation gradient tensor that are conditioned upon the Lagrangian kinematics~\citep{Lester:2018aa,Lester:2021aa}. In turn, these stochastic models for fluid deformation models can be applied to studies of colloidal transport, chemical reactivity, solute mixing and dispersion, and can be used as a based for the design of novel porous materials with tuneable transport properties.

Dynamical system methods have also provided a connection between the pore-scale architecture and the Lyapunov exponents in random and ordered porous media. For ordered porous media, \citet{Turuban:2018aa,Turuban:2019aa} analyse symmetry breaking of the porous architecture with respect to the mean flow direction to develop predictive models of fluid stretching. Similarly, \cite{Heyman:2020aa,Heyman:2021aa} use dynamical system methods to develop predictive models of fluid stretching for random granular media based upon mean coordination number of the granular assembly. These dynamical systems approaches provides a systematic consistent methodology to connect the Lagrangian kinematics of pore-scale flows with the underlying porous architecture and quantify their impact upon important processes such as solute mixing and dispersion.

\subsection{Darcy Scale --- Subsurface Remediation}

Conventional methods to remediate or treat groundwater contaminants have typically relied on simple ``pump and treat'' or \emph{in situ} reactant injection interventions to reduce contaminant concentrations, often with a steady or quasi-steady deployment schedule and augmented by the emplacement \emph{in situ} of permeability barriers which seek to control fluid migration so that contaminants are more easily targeted for extraction or for \emph{in situ} chemical treatments.

Dynamical systems approaches provide a different tool set to address such problems. Many subsurface groundwater environments are well approximated by 2D domains so it is easy to see from Figure \ref{fig:Darcy_chaos_diagram} that the addition of time-dependent forcing will be sufficient to admit the possibility of LCS in the subsurface flow field. Using a symmetric arrangement of pumping wells led to the concept of Rotated Potential Mixing (RPM) flows which were shown theoretically, computationally and experimentally to provide a ready means to control fluid mixing and segregation in porous media at scales ranging from centimetre to metre scales \citep{Metcalfe:2010ab,Metcalfe:2010aa,Lester:2009ab,Lester:2010aa,Trefry:2012ac,Trefry:2012aa,Rodriguez:2017aa} and then at field scale at the Borden site where the RPM schedule was varied in order to promote chaotic mixing of injected tracer \citep{Cho:2019aa}. This kind of programmed pumping scheme is also known in the groundwater literature as ``engineered injection and extraction'' (see, e.g. \citet{Wang:2022aa}).

Similar thinking has led to the elaboration of so-called ``active spreading'' engineering techniques \citep{Sather:2022aa} which utilize flow field variations to enlarge the reaction front between subsurface reactants. In this approach detailed exposition of LCS is not pursued in favour of relying on the enhanced mixing performance of transient flows in heterogeneous formations \citep{Mays:2012,Piscopo:2013aa,Neupauer:2015aa} that is guaranteed according to Figure \ref{fig:Darcy_chaos_diagram}. Thus the dynamical systems approach underpins the recognition that (1) transient forcing in heterogeneous porous media will inherently improve fluid mixing, and that (2) the improvement can be both quantified and controlled using the tools of Lagrangian kinematics. Apart from this environmental engineering context, this knowledge has likely beneficial potential in  the mining industry (e.g. \emph{in situ} mining, mineral processing) and in geothermics (e.g. ground source heat pumps, heat reservoir management), as well as in many other applications.

\section{Conclusions}
\label{sec:conclude}

Complex transport dynamics (such as chaotic advection) abounds in porous media flow. It is inherent to almost all pore-scale flows, and arises in many scenarios at the Darcy and regional scales. Complex Lagrangian kinematics differ from conventional mixing dynamics in that fluid deformation may be exponential in time, and the underlying dynamics often cannot be fully understood in the Eulerian frame. More than just a quantitative change, exponential fluid stretching profoundly augments solute mixing and transport, chemical reactions and biological activity in porous media. Specifically, unlike sub-exponential fluid stretching, exponential fluid stretching leads to \emph{singular} solute mixing~\citep{Cerbelli:2017aa} in that the limit of vanishing diffusivity does not correspond (qualitatively or quantitatively) to the zero diffusivity case. Hence the impacts of chaotic advection are typically not captured by conventional methods of modelling transport in porous media, and so unfamiliar tools and techniques are required to visualise, understand, quantify and, in engineering applications, ultimately control and exploit this phenomenon.

Hitherto the dynamical systems approach has not been widely appreciated within the community of practitioners of flow and transport in porous media, and this article unabashedly hopes to address that. This is not to say that chaotic advection, or more precisely transport mediated by Lagrangian coherent structures, is a magic bullet that solves all outstanding problems the field. On the contrary, we merely state that students and practitioners should be aware of this transport modality when circumstances warrant, and include appropriate techniques amongst the arsenal of conventional methods (stochastic methods, upscaling techniques, uncertainty analysis) to understand and analyse transport in porous media.  However, these circumstances are not rare, rather they are common in porous media flows. As chaotic dynamics are the norm rather than exception in classical systems, chaos arises in most situations except for those where kinematic or topological constraints explicitly forbid it. For historical reasons the mental picture cultivated for porous media transport excluded complex Lagrangian kinematics, but now the dynamical systems approach to transport should be in your toolkit.

One important conceptual tool is the advective template of the flow which is given by the evolution of non-diffusive fluid tracer particles, and so explicitly resolves the system Lagrangian kinematics, associated coherent structures and topology. For any particular application of porous media flow, this advective template necessarily provides an incomplete description of the system as almost always there will be additional physical phenomena such as dispersion, diffusion, chemical  reactions and biological activity, all of which interact nonlinearly with the advective template.  We argue, however, that for every application knowledge of the underlying advective template is in and of itself useful and informative as this forms the organising structure upon which all additional fluid-borne phenomena play out. It is for this reason that we have focused this primer mainly on the advective template, yet nearly everywhere else in the porous media flow literature it is neglected. We note that many additional tools and techniques applicable to porous media can be found in the dynamical systems (chaos) literature in general and chaotic advection in particular.

Although deliberately limited in scope, several key messages are conveyed in this primer. First, chaotic advection is common to many porous media applications across all length scales. Second, these complex Lagrangian kinematics qualitatively augment associated fluid-borne phenomena, neither of which can be fully understood without utilisation of appropriate tools and techniques. Third, there exist easily identifiable kinematic constraints in particular applications that explicitly preclude chaotic advection, and chaos is expected when these constraints are absent. Fourth, fundamental studies of chaotic advection point toward development of novel classes of porous hierarchical materials (from pore to Darcy scales) with tuneable transport properties. Fifth, saturated porous media may now be regarded as a reaction vessel whose mixing state can be manipulated to promote reactivity or segregation according to purpose. Finally, many natural systems exhibit chaotic advection and complex Lagrangian transport due to the unsteady forcings inherent to these systems. 

A new way of looking at a problem opens up myriad new and interesting research questions. Key knowledge gaps in the field include: (i) development and application of methods for characterisation of chaotic advection in all real (i.e. not model) natural and engineered porous materials, (ii) how the kinematics of steady 3D flow at the pore-scale influence macroscopic fluid mixing and chemical transport in saturated porous media, (iii) how to develop upscaled models of dispersion and dilution which capture these processes, (iv) design rules for engineered porous materials with tuneable transport properties, (v) development of topologically consistent upscaling methods for Darcy flows that do not introduce spurious kinematics, (vi) merging of chaotic advection methods with stochastic methods and conventional approaches to transport in porous media. This topological approach to fluid physics at the pore scale is a relatively new area and opportunities abound for further development. Another knowledge gap is how these concepts may be extended to real porous architectures via the development of stochastic models for the evolution of the fluid deformation tensor conditioned on tomographic imaging of the pore space and computational fluid dynamics simulations. Finally, more experiments are needed looking for the effects of chaotic advection across porous media applications.

Our goal with this paper is to implant in the reader the \textit{chaotic} way of thinking and to generate some familiarity with the requisite mathematical apparatus. By this route we hope to open up new ways of designing porous media reaction technology and new understandings of ecological and biological porous transport. Let us know if we succeed.

\begin{acknowledgements}
GM thanks the Lorenz Center for travel support and hospitality. Thoughtful and constructive comments of two reviewers helped improve this manuscript.
\end{acknowledgements}

\bibliographystyle{spbasic}      
\bibliography{database.refLorentz.bib}   


\section{Statements and  Declarations}

The authors declare that no funds, grants, or other support were received during the preparation of this manuscript. The authors have no relevant financial or non-financial interests to disclose. The manuscript was written and drafted by all authors with equivalent contributions. All authors read and approved the final manuscript. No data sets were generated or analysed during the current study.

\end{document}